\documentclass[aps,pra,twocolumn,superscriptaddress, nofootinbib]{revtex4-1}

\usepackage{amsmath}
\usepackage{amssymb}
\usepackage{mathrsfs}
\usepackage{verbatim}
\usepackage{epsfig}
\usepackage{color}
\usepackage{soul}
\usepackage[normalem]{ulem}
\usepackage[bookmarks]{hyperref}
\usepackage{orcidlink}
\usepackage{graphicx}
\usepackage{derivative}
\usepackage{subcaption}
\usepackage{aligned-overset}
\usepackage{nicefrac}
\usepackage{braket}
\usepackage{tcolorbox}
\graphicspath{{.}{./EPS/}}

\begin{document}

\title{Analytical solutions for optimal photon absorption into inhomogeneous spin memories}
%\title{Optimized photon absorption in cavity-coupled inhomogeneous spin ensembles}

\author{J\'ozsef Zsolt Bern\'ad}
\email{j.bernad@fz-juelich.de}
\affiliation{Forschungszentrum Jülich, Institute of Quantum Control,
Peter Grünberg Institut (PGI-8), 52425 Jülich, Germany}

\author{Michael Schilling}
\affiliation{Forschungszentrum Jülich, Institute of Quantum Control,
Peter Grünberg Institut (PGI-8), 52425 Jülich, Germany}
\affiliation{Institute for Theoretical Physics, University of Cologne, 50937 Köln, Germany}

\author{Yutian Wen}
\affiliation{Universit\'{e} Paris-Saclay, CEA, CNRS, SPEC, 91191 Gif-sur-Yvette Cedex, France}

\author{Matthias M. Müller}
\affiliation{Forschungszentrum Jülich, Institute of Quantum Control,
Peter Grünberg Institut (PGI-8), 52425 Jülich, Germany}

\author{Tommaso Calarco \orcidlink{0000-0001-5364-7316}}
\affiliation{Forschungszentrum Jülich, Institute of Quantum Control,
Peter Grünberg Institut (PGI-8), 52425 Jülich, Germany}
\affiliation{Institute for Theoretical Physics, University of Cologne, 50937 Köln, Germany}
\affiliation{Dipartimento di Fisica e Astronomia, Universit\'{a} di Bologna, 40127 Bologna, Italy}

\author{Patrice Bertet}
\affiliation{Universit\'{e} Paris-Saclay, CEA, CNRS, SPEC, 91191 Gif-sur-Yvette Cedex, France}

\author{Felix Motzoi \orcidlink{0000-0003-4756-5976}}
\affiliation{Forschungszentrum Jülich, Institute of Quantum Control,
Peter Grünberg Institut (PGI-8), 52425 Jülich, Germany}

\date{\today}

\begin{abstract}
We investigate for optimal photon absorption a quantum electrodynamical model of an inhomogeneously-broadened spin ensemble coupled to a single-mode cavity. Solutions to this problem under experimental assumptions are developed in the Schr\"odinger picture without using perturbation theory concerning the cavity-spin interactions.
Furthermore, we exploit the possibility of modulating the frequency and coupling rate of the resonator. We consider a one-photon input pulse and show some optimal scenarios, where exact formulas and numerical results are obtained for the absorption probabilities and the optimal pulse shapes. In particular, if the external loss dominates the internal loss of the cavity, we find the optimal cooperativity for different parameters and identify cases where absorption with a success probability larger than $99\%$ is achieved. 
\end{abstract}

\maketitle

\section{Introduction}
\label{I}

Information transport is fundamental to the scalability of both short and long-range quantum architectures. A chief candidate is the flying mode approach where information carriers are themselves quantum objects that must retain coherence over the intermediary channel.  Applications of quantum transport range widely, including quantum communication \cite{Sangouard}, remote sensing \cite{schnabel2010quantum}, optical computation \cite{kok2007linear,  slussarenko2019photonic,cohen2018deterministic}, error correction \cite{kerckhoff2010designing, martin2015deterministic}, and cryptography \cite{pirandola2020advances}. For quantum computing purposes, flying modes are key to scalability, e.g.~via traveling electrons \cite{fujita2017coherent,li2018crossbar}, ions \cite{kaushal2020shuttling}, atoms \cite{bluvstein2022quantum}, or photons \cite{roch2014observation, reiserer2014quantum, Rempe2020} between static qubits. They enable larger spacings, better connectivity, and connection to storage qubits with longer lifetimes \cite{grezes2014multimode}. In this context, we will study a situation where a quantum memory is composed of multiple inhomogeneously cavity-coupled and broadened matter qubits. This can be formed with solid-state electronic spins  \cite{Morley2010, Pohl2012,grezes2014multimode,Afzelius2018, Sullivan2020, Bertet2020,Morton2022, Wen2022}.

In the last two decades, many different   approaches towards a working quantum memory have been proposed. The storage and retrieval of single photons was achieved by employing electromagnetically
induced transparency combined with the Duan-Lukin-Cirac-Zoller protocol \cite{EIT1, EIT2, EIT3}. However, to enable a long-lived quantum memory, the community has started to explore solid-state materials, which have long coherence lifetimes and large inhomogeneous widths. In these approaches, rephasing pulses are used and therefore enormous  population inversion in the material is produced. To avoid that, a considerable literature has arisen, where protocols like controlled reversible inhomogeneous broadening \cite{Moiseev2001, Nilsson2005, Kraus2006, Gorshkov2007, Lauritzen2010, Wu2010}, the gradient echo memory \cite{Hetet1, Longdell2008, Hetet2, Hedges2010}, and the atomic frequency comb \cite{deRiedmatten2008, Afzelius2009} have been investigated. The other alternative is to kill the primary echo and the corresponding scheme of the revival of silenced echo was experimentally also demonstrated \cite{McAuslan2011, Damon2011}. In this work, we are motivated to optimise the very first step of all these techniques, namely the absorption of the flying photon by the spin ensemble. We analytically derive absorption equations and numerically optimise the external drive to minimise the time and maximise the efficiency.
The efficient storage of a photon into a single atom has already been investigated both experimentally \cite{Rempe2020} and theoretically \cite{Alber2016}. However, in the case of a spin ensemble, each spin is characterised by an individual magnetic dipole coupling and its Larmor frequency within an inhomogeneous 
linewidth. The mathematical description of a similar system was introduced almost three decades ago \cite{Garraway96, Garraway97}, where the authors  treated the spontaneous decay of $V$-type atomic systems into different photonic band gaps. Here, a single-mode of the radiation field interacts with spins which results in a spin-induced cavity linewidth \cite{Staudt}, i.e., the cavity field decays into the spin ensemble. This model is usually investigated under semi-classical approximations or by employing perturbation techniques \cite{Kurucz, Diniz, Brian2012, Afzelius, Bertet2013}. These approaches focus on the Heisenberg picture, where expectation values of the system's operators are calculated, and this leads to a system of infinitely many differential equations. Our focus lies on the Schr\"odinger picture, where the solution to the time evolution of the state becomes tractable due to the presence of a single excitation.

In this work, we have three aims: first, to present a minimal model with a spin ensemble and a cavity, which can describe the storage process of an 
incoming photon with tunable decay rate and detuning of the cavity \cite{Wen2022}; second, to non-perturbatively describe the time evolution of this model with arbitrary input waveforms and inhomogeneous broadening distributions, including Lorentzians and Gaussians; third, to probe the optimality of the storage with simple few-parameter pulses. In particular, we investigate both closed-form and numerically optimised solutions, both for sequential absorption of the photon into the cavity followed by the spin ensemble, or directly from the external field into the ensemble.     

The paper is organised as follows. In Sec.~\ref{II}, we set notation, introduce the Hamiltonian model describing the external modes of the radiation field, and derive the input-output theory for ensemble spins in Schr\"odinger picture. In Sec.~ \ref{III}, we derive the exact solution to the complete time evolution of the system. In Sec.~\ref{IV}, we analyse the two-step sequential excitation of the cavity and spin ensemble, which is made possible by the controllability of the cavity parameters, and find a globally optimal protocol.  In Sec.~\ref{V}, we investigate a direct, single-step approach to the excitation of the spin ensemble. Further optimisations using both numerical and analytical approaches are presented. In Sec.~\ref{VI},  we summarise our conclusions. Details supporting the main text are collected in the Appendix.  

\section{Model}
\label{II}

In this section, we develop the system model for the inhomogeneous spin ensemble system coupled to a microwave cavity and fix the required notation. In this particular system, experimental studies have already demonstrated either in rare-earth ion-doped crystals \cite{Afzelius2018, Wen2022} or for bismuth donors in silicon \cite{Morley2010,Sullivan2020} that the relaxation times $T_1$ and $T_2$ are larger than a few milliseconds with $T_1>T_2$. A recent study has even shown that a bismuth-doped silicon, which was coupled to an aluminum resonator, has a $T_2$ time as large as $0.3$ s \cite{Bertet2020}. Motivated by these experimental results and considering up to $100 \mu$s long absorption processes, we conclude that decoherence and dissipation effects on the spins can be excluded from our model. In general, the spins are not isolated from the environment, which, in our case, consists of modes of the radiation field and also material degrees of freedom. As we have $N$ nonidentical spins and they share the same environment, it is immediate that spin-spin interactions and further collective effects are induced through the environment \cite{Tanas,Carnio}. Our main aim is to obtain optimal absorption of a photon, which takes place on a much faster time scale, however, seeking further tasks like a few hundred milliseconds storage and then retrieval, when these environment-induced effects ought to be modelled as well. In the subsequent discussion, based on the above arguments, we present our model.

\begin{figure}[t!]
 \includegraphics[trim=21cm 2.5cm 1cm 4cm, clip=true, width=0.5\textwidth]{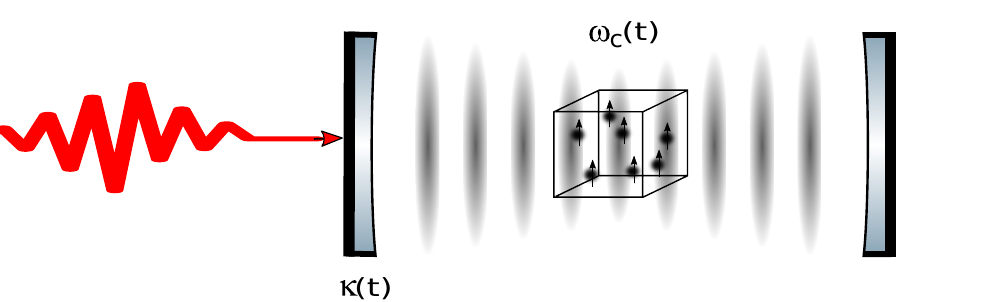}
\caption{Schematic representation of the cavity QED scenario. The frequency $\omega_c$ and the total decay rate $\kappa$ of the single-mode cavity field are tunable. An external driving field is applied to the spin ensemble via the cavity. The aim of the control problem is the optimal absorption of an incoming photon by the spin ensemble.}
   \label{fig:setup}
\end{figure}

Let us consider a single cavity mode interacting with external and internal radiation fields and an ensemble of $N$ spins within the cavity as schematically depicted in Fig. \ref{fig:setup}. The single-mode is supported by the cavity, while the internal radiation field refers to the remaining modes in the cavity. We assume that the 
dipole and rotating-wave approximations are valid for this setup. The $i$th spin system comprises a ground state 
level $\ket{0}_i$ and an excited level $\ket{1}_i$ of different parity and they are separated by an energy difference $\hbar \omega_i$. It is assumed that the frequency $\omega_i$ is inhomogeneously broadened 
around a central frequency $\omega_s$. The magnetic dipole coupling $g_i$ of each spin, which involves the transition dipole moment of the states 
and the normalised mode function of the single-mode radiation field in the cavity, is distributed also inhomogeneously. A microscopic model of the external radiation field allows us to describe the 
propagation of photons outside the cavity and photons can enter this cavity by transmission through a mirror. We also assume that the typical interaction times are small enough to neglect the spontaneous 
emission of photons from any excited state $\ket{1}_i$.  Within these considerations, the evolution is described by the Hamiltonian
\begin{eqnarray}
 \hat{H}/\hbar&=&\omega_c \hat{a}^\dagger \hat{a}+\sum^N_{i=1} \frac{\omega_i}{2} \hat{\sigma}^{(i)}_z + \sum^N_{i=1} g_i \left(\hat{a} \hat{\sigma}^{(i)}_+ +\hat{a}^\dagger \hat{\sigma}^{(i)}_- \right) \nonumber \\
 &+& \sum_{j\in L} \Omega_j \hat{a}^\dagger_j\hat{a}_j + 
\sum_{j \in L} \left( \kappa_j 
\hat{a}^\dagger_j \hat{a}+ \kappa^*_j 
\hat{a}^\dagger \hat{a}_j\right), \label{eq:hamiltonian_first} \\
 &+& \sum_{k \in K} \tilde{\Omega}_k \hat{b}^\dagger_k\hat{b}_k + 
\sum_{k \in K} \left( \tilde{\kappa}_k 
\hat{b}^\dagger_k \hat{a}+ \tilde{\kappa}^*_k 
\hat{a}^\dagger \hat{b}_k\right), \nonumber
\end{eqnarray}
where $\hat{\sigma}^{(i)}_z=\ket{1}\bra{1}_i-\ket{0}\bra{0}_i$, $\hat{\sigma}^{(i)}_+=\ket{1}\bra{0}_i$, and $\hat{\sigma}^{(i)}_-=\ket{0}\bra{1}_i$. The annihilation and creation
operators of the single-mode radiation field in the cavity with frequency $\omega_c$ are denoted by $\hat{a}$ and $\hat{a}^\dagger$. The external field is considered to have a set of modes, which is denoted by $L$,
and $\hat{a}_j$ ($\hat{a}^\dagger_j$) is the annihilation (creation) operator of the $j$th mode with frequency $\Omega_j$. The coupling $\kappa_j$ gives the interaction strength between the 
single-mode of the cavity and the $j$th mode of the external field. There is an additional effect, the internal cavity loss, which is modelled by the interaction of the single-mode field with the other modes of the cavity denoted by the index set $K$, where $\hat{b}_k$ ($\hat{b}^\dagger_k$) is the annihilation (creation) operator of the $k$th mode with frequency $\tilde{\Omega}_k$. Usually, this occurs, because the photon is scattered by the mirrors out of the mode and the strength of this effect is given by $\tilde{\kappa}_k$ for the $k$th mode of the internal field. 
%%%%%%%%%%%%%%%%%%%%%%%%%%%%%%%%%%%%
%%%%%%%%%%%%%%%%%%%%%%%%%%%%%%%%%%%%

In order to describe the dynamics in the rotating frame of the central spin frequency $\omega_s=\sum^N_{i=1}\omega_i/N$, we use the unitary transformation $\hat U(t)=e^{i \hat H_0 t/\hbar}$ with 
\begin{equation}
 \hat H_0/\hbar=\omega_s \hat{a}^\dagger \hat{a}+\sum^N_{i=1} \omega_s\hat{\sigma}^{(i)}_z/2 + \sum_{j\in L} \omega_s \hat{a}^\dagger_j\hat{a}_j + \sum_{k\in K} \omega_s \hat{b}^\dagger_k\hat{b}_k    
\end{equation}
 to obtain the transformed state vector. It follows that the transformed Hamiltonian is
\begin{equation}
 \hat{H}_{\text{I}}=\hat U(t) \hat{H} \hat{U}^\dagger(t)+i\hbar \dot{\hat{U}}(t) \hat{U}^\dagger(t),    
\end{equation}
which reads 
\begin{eqnarray}
 \hat{H}_{\text{I}}/\hbar&=&\Delta_{cs} \hat{a}^\dagger \hat{a}+\sum^N_{i=1} \frac{\Delta_i}{2} \hat{\sigma}^{(i)}_z + \sum^N_{i=1} g_i \left(\hat{a} \hat{\sigma}^{(i)}_+ +\hat{a}^\dagger \hat{\sigma}^{(i)}_- \right) \nonumber \\
 &+& \sum_{j\in L} \delta_j \hat{a}^\dagger_j\hat{a}_j + 
\sum_{j \in L} \left( \kappa_j 
\hat{a}^\dagger_j \hat{a}+ \kappa^*_j 
\hat{a}^\dagger \hat{a}_j\right), \label{eq:Hamiltonian} \\
&+& \sum_{k\in K} \tilde{\delta}_k \hat{b}^\dagger_k\hat{b}_k + 
\sum_{k \in K} \left( \tilde{\kappa}_k 
\hat{b}^\dagger_k \hat{a}+ \tilde{\kappa}^*_k 
\hat{a}^\dagger \hat{b}_k\right), \nonumber
\end{eqnarray}
where $\Delta_{cs}=\omega_c-\omega_s$, $\Delta_i=\omega_i-\omega_s$, $\delta_j=\Omega_j-\omega_s$, and $\tilde{\delta}_k=\tilde{\Omega}_k-\omega_s$. It is easy to check that this interaction Hamiltonian commutes with the particle number operator
\begin{equation}
 \hat{N}= \hat{a}^\dagger \hat{a} + \sum^N_{i=1} \hat{\sigma}^{(i)}_+ \hat{\sigma}^{(i)}_- +  \sum_{j\in L} \hat{a}^\dagger_j\hat{a}_j + \sum_{k\in K} \hat{b}^\dagger_k\hat{b}_k, \nonumber
\end{equation}
and thus $\hat{N}$ is a conserved quantity. We assume only one excitation in the whole system, which yields that the form of the state vector
\begin{eqnarray}
 \ket{\Psi}&=& \sum^N_{i=1} \Psi^{(i)}_s |0\dots \underbrace{1}_{i\text{th}\,\, \text{position}} \dots 0\rangle_s \ket{0}_c \ket{0}_E \ket{0}_I \nonumber \\
 &+& \sum_{j \in L} \Psi^{(j)}_E \ket{0}_s \ket{0}_c |0\dots \underbrace{1}_{j\text{th}\,\, \text{position}} \dots 0\rangle_E \ket{0}_I \nonumber \\
 &+& \sum_{k \in K} \Psi^{(k)}_I \ket{0}_s \ket{0}_c \ket{0}_E |0\dots \underbrace{1}_{k\text{th}\,\, \text{position}} \dots 0\rangle_I \nonumber \\
 &+& \Psi_c \ket{0}_s \ket{1}_c \ket{0}_E \ket{0}_I, \label{eq:ansatz}
\end{eqnarray}
is preserved during the time evolution. Here $\Psi^{(i)}_s$, $\Psi^{(j)}_E$, $\Psi^{(k)}_I$ and $\Psi_c$ denote the probability amplitudes to find an excitation in the $i$th spin, $j$th mode of the external field, $k$th mode of the internal field, and the mode
of the cavity. Furthermore, we have used the following simplified notations: $\ket{0}_s=\ket{000\dots0}_s$, i.e., all spin are in the ground state, $\ket{0}_E=\ket{000\dots0}_E$, i.e., there is no 
photon in the modes of the external field, and $\ket{0}_I=\ket{000\dots0}_I$, i.e., there is no 
photon in the other modes of the internal field. 
This allows the derivation of the following equations of motion
\begin{eqnarray}
  \dot{\Psi}^{(i)}_s &=&i (\Delta_N -\Delta_i)  \Psi^{(i)}_s -i g_i \Psi_c, \label{eq:s} \\   
  \dot{\Psi}_c&=&-i  (\Delta_{cs}- \Delta_N) \Psi_c -i \sum^N_{i=1} g_i \Psi^{(i)}_s \nonumber \\ 
  &-&i \sum_{j \in L} \kappa^*_j \Psi^{(j)}_E - i \sum_{k \in K} \tilde{\kappa}^*_k \Psi^{(k)}_I,  \label{eq:c} \\
  \dot{\Psi}^{(j)}_E&=&- i (\delta_j-\Delta_N) \Psi^{(j)}_E-i\kappa_j \Psi_c, \label{eq:E}, \\
\dot{\Psi}^{(k)}_I&=&- i (\tilde{\delta}_k-\Delta_N) \Psi^{(k)}_I-i\tilde{\kappa}_k \Psi_c, \label{eq:I} 
\end{eqnarray}
with
\begin{equation}
 \Delta_N=\sum^N_{i=1} \frac{\Delta_i}{2}. \nonumber
\end{equation}
The solution to these coupled equations is complicated by the presence of finite but large and countable infinite summations. For the initial conditions we assume that
initially the excitation is in the external field
\begin{eqnarray}
 &&\Psi^{(i)}_s(0)=0, \quad \Psi_c (0)=0, \quad \sum_{j \in L} |\Psi^{(j)}_E(0)|^2=1, \nonumber  \\
 && \text{and} \quad \sum_{k \in K} |\Psi^{(k)}_I(0)|^2=0. \label{eq:in}
\end{eqnarray}
To get a system of differential equations that does not involve explicitly the external and the internal field, we first integrate \eqref{eq:E},
\begin{eqnarray}
\Psi^{(j)}_E(t)&=& \Psi^{(j)}_E(0) e^{-i (\delta_j-\Delta_N) t} \label{eq:intE}\\
&-&i\kappa_j \int^t_0 e^{-i (\delta_j-\Delta_N) (t-t')} \Psi_c(t')\, dt' \nonumber
\end{eqnarray}
and then \eqref{eq:I}
\begin{eqnarray}
\Psi^{(k)}_I(t)=-i\tilde{\kappa}_k \int^t_0 e^{-i (\tilde{\delta}_k-\Delta_N) (t-t')} \Psi_c(t')\, dt' \label{eq:intI}.
\end{eqnarray}

On substituting these expressions into \eqref{eq:c}, we obtain
\begin{widetext}
\begin{eqnarray}
 \dot{\Psi}_c&=&-i  (\Delta_{cs}- \Delta_N) \Psi_c -i \sum^N_{i=1} g_i \Psi^{(i)}_s-i \sum_{j \in L} \kappa^*_j \Psi^{(j)}_E(0) e^{-i (\delta_j-\Delta_N) t} - \sum_{j \in L} |\kappa_j|^2 
 \int^t_0 e^{-i (\delta_j-\Delta_N) (t-t')} \Psi_c(t')\, dt'  \nonumber \\
 &-&  \sum_{k \in K} |\tilde{\kappa}_k|^2 
 \int^t_0 e^{-i (\tilde{\delta}_k-\Delta_N) (t-t')} \Psi_c(t')\, dt'. \label{eq:cone}
\end{eqnarray}
\end{widetext}
Assuming that the quantization volume $V$ is very large then the modes of the external field will be closely spaced in frequency. Therefore, we can replace the summation over the modes with an integral:
\begin{equation}
\label{eq:sumtoint}
\sum_{j \in L} \to n_p \int_{V} d^3 \mathbf{k}\, \rho_E(\mathbf{k}),
\end{equation}
where $\rho_E(\mathbf{k})$ is the density of states and $\mathbf{k}$ the wave vector. The factor $n_p$ accounts for the possible polarization directions for each mode of the external field. In general, there are two polarization directions, but in certain cases, $n_p$ is equal to one \cite{Schleich}. We furthermore assume, based on the phenomenology of such cavities \cite{Wen2022}, that the scattering of the photon out of the supported mode into the other internal modes can be modelled as a Markovian loss process. However, this time the quantisation volume, i.e., the volume of the cavity, is finite and thus we have countable infinite modes. In the phenomenological modeling to obtain a Markovian loss process, we consider the couplings $\tilde{\kappa}_k$ to be such that the time on which the transferred energy is fed back to the cavity mode is practically very long on the time scale of the dynamic  \cite{Weiss}. Then, we may replace the sum over $K$ by an integral over a continuous spectral density. This results in an effective model and the Weisskopf-Wigner approach \cite{Weisskopf1, Weisskopf2} for our case yields  
\begin{eqnarray}
 \dot{\Psi}_c&=&-i  (\Delta_{cs}- \Delta_N) \Psi_c -i \sum^N_{i=1} g_i \Psi^{(i)}_s+f_e(t) \nonumber \\
 &-& \frac{\kappa_e+\kappa_i}{2}\Psi_c,  \label{eq:ctwo} 
\end{eqnarray}
where $\kappa_e$ and $\kappa_i$ are the external and internal decay rates of the cavity. We introduce now the total decay rate $\kappa=\kappa_e+\kappa_i$. It is usually assumed that the radiation in the external field is going to be centred about the single-mode cavity wave number $k_c=\omega_c/c$, where $c$ is the speed of light. In the limit of continuum modes of the external field, $|\kappa_j|^2$ in Eq.~\eqref{eq:cone} is replaced by $|\kappa_e (\mathbf{k})|^2$, which has the dimensionality of $1/\mathrm{s}^2$ and is approximated as $|\kappa_e (k_c \mathbf{e}_k)|^2$ with $\mathbf{e}_k=\mathbf{k}/|\mathbf{k}|$. The integration over time in Eq.~\eqref{eq:cone} fixes the dimensionality of $\kappa_e$, i.e., the decay rate has frequency 
dimension. Furthermore, we have
\begin{equation}
 f_e(t)=-i n_p \int_{V} d^3 \mathbf{k}\, \rho_E(\mathbf{k}) \kappa^*_e (\mathbf{k}) \Psi_E(\mathbf{k}, 0) e^{-i (c |\mathbf{k}|-\omega_s-\Delta_N) t}. \label{eq:initial}
\end{equation}
The external drive $f_e(t)$ of the probability amplitude $\Psi_c$ depends on $\Psi^{(j)}_E(0)$,
$\kappa_j$s, and all the frequencies introduced within this model. It is worth noting
that the external field depends not only
on $\Psi^{(j)}_E(0)$ but also on spatial coordinates via the orthonormal mode functions. These functions are solutions to the Helmholtz equation, fulfil the boundary conditions of the volume $V$, and satisfy the Coulomb gauge condition. 
The external drive $f_e(t)$ is subject to the constraint 
\begin{equation}
   \int^\infty_{-\infty} |f_e(t)|^2 \,dt=\kappa_e ,\label{eq:constraint}
\end{equation}
which we derive in Appendix \ref{AppendixB}. Beyond this, $\dot{\Psi}_c$ in Eq.~\eqref{eq:ctwo} only explicitly depends on $\kappa$ not $\kappa_e$. We use this to represent the following results as rescaled versions of the $\kappa_i=0$ case by defining the rescaling parameter $\alpha=\sqrt{\frac{\kappa_e}{\kappa}}$ and $\int^\infty_{-\infty} |f(t)|^2 \,dt=\kappa$, so that $f_e(t) = f(t) \alpha$.\\

Finally, it is worth noting that our motivation to obtain solutions also for the external field in Eq. \eqref{eq:intE}, is similar to the input-output theory of Collett and Gardiner \cite{Collett}. They derive the Heisenberg equations of motion without any two-level systems inside the cavity and introduce the input and output operators. We have a spin ensemble in the cavity and employ the Schrödinger equation, which is possible due to the presence of a single photon.\\

\section{Exact time-dependent solution}\label{III}

Given the equations above, we show how it can be formally solved for a given shape of the input field. To do so, we integrate \eqref{eq:s},
\begin{equation}
 \Psi^{(i)}_s=-i g_i \int^t_0 e^{i (\Delta_N -\Delta_i) (t-t')} \Psi_c(t')\,dt', \label{eq:psis-integrodifferential}
\end{equation}
and on substituting this expression into \eqref{eq:ctwo} we obtain
\begin{eqnarray}
 \dot{\Psi}_c&=&-i  (\Delta_{cs}- \Delta_N) \Psi_c +\alpha f(t) - \frac{\kappa}{2}\Psi_c \nonumber \\
 &-&\sum^N_{i=1} g^2_i \int^t_0  e^{i (\Delta_N -\Delta_i) (t-t')} \Psi_c(t')\,dt'. \label{eq:cthree}
\end{eqnarray}
It is worth noting that we have replaced many linear differential equations with one linear integro-differential equation. Next, we replace the summation over the spins with an integral by considering

In this paper, we use as our base case a Lorentzian-broadened spin ensemble
\begin{equation}
 p_1(\Delta)=\frac{w}{2 \pi}  \frac{1}{\Delta^2+w^2/4}, \label{eq:broaden}
\end{equation}
where $w$ is the linewidth or broadening. Other distributions are discussed in section~\ref{sec:other_distributions}. Then, we have
\begin{eqnarray}
 \Delta_N=\sum^N_{i=1} \frac{\Delta_i}{2} \rightarrow \int^\infty_{-\infty} p_1(\Delta) \frac{\Delta}{2}\, d\Delta \label{eq:DN}
\end{eqnarray}
and the Cauchy principal value of the integral is zero, because $p_1(\Delta) \Delta$ is an odd function. Thus, $\Delta_N=0$. In the case of the coupling strengths, we have
\begin{equation}
 \sum^N_{i=1} g^2_i \rightarrow \int^\infty_{-\infty} p_2(g) g^2 \, dg =  g^2_{\text{ens}}, \label{eq:inhomog}
\end{equation}
where $g_{\text{ens}}$ is the ensemble-coupling constant. The coupling-strength distribution function $p_2(g)$ is determined by experimental measurements. Finally, we use again the Cauchy principal value theorem
to obtain
\begin{widetext}
\begin{equation}
 \sum^N_{i=1} g^2_i e^{i (\Delta_N -\Delta_i) (t-t')} \rightarrow \int^\infty_{-\infty}\, d\Delta \int^\infty_{-\infty}\, d g
 p_1(\Delta)  p_2(g) g^2 e^{-i \Delta (t-t')}= g^2_{\text{ens}} e^{-w (t-t')/2}.
\end{equation}
\end{widetext}
Hence, Eq.~\eqref{eq:cthree} reads
\begin{equation}
 \dot{\Psi}_c=-i \Delta_{cs} \Psi_c-g^2_{\text{ens}} \int^t_0  e^{-w (t-t')/2} \Psi_c(t')\,dt'+\alpha f(t) - \frac{\kappa}{2}\Psi_c. \label{eq:cfinal}
\end{equation}
Eq.~\eqref{eq:cfinal} together with \eqref{eq:s} and \eqref{eq:E} yields a complete description of the system's evolution. Our choice of the Cauchy-Lorentz distribution in \eqref{eq:broaden} is motivated by the 
fact that the characteristic function of this probability distribution has a Laplace transform involving only one polynomial, which plays an essential role, when the inverse Laplace transformation is applied. There are other 
probability distributions \cite{Abramowitz}, which fulfil this mathematical requirement, and if it is required, their convex combinations can be used to define an 
experimentally more suitable $p_1(\Delta)$. A counterexample is the Gaussian distribution because its characteristic function is also a Gaussian function and its Laplace transform involves the error function (see Appendix \ref{AppendixA}).

Our aim in the subsequent sections is to describe the optimal storage of the incoming single photon in the spin ensemble. To this end, we use the tunability of $\omega_c$ and $\kappa$. We will investigate the dynamics of the model of Sec. \ref{II} to obtain those conditions, which allow the optimal storage of a photon in the spin ensemble. We remind the reader of the discussion at the beginning of Sec. \ref{II} that the population and phase decays of each spin are neglected due to their longer characteristic times than the absorption process. Finally, these types of systems are subject to large values of broadening, i.e., $w/2\pi$ can be $10$ MHz \cite{Wen2022}, while $g_{\text{ens}}/2 \pi$ and $\kappa/2\pi$ are always smaller than $1$ MHz. Therefore, without the loss of generality, throughout the whole paper we are going to use the following condition:
\begin{equation}
 \frac{w}{\kappa}, \frac{w}{g_{\text{ens}}}>5. \label{eq:imcond}    
\end{equation}
This implies that Eq.\eqref{eq:cfinal} will describe two decay phenomena, the photon either leaves the cavity or gets absorbed by the spin ensemble. In this situation, there will be no emission from the spin ensemble, unless a refocusing of the spins is performed. The retrieval of the photon will be not discussed in this work, but we seek optimal strategies to deposit it into the spin ensemble under the above-presented circumstances. We will analyze two protocols: first, we will study a two-step protocol, where the photon is first brought into the cavity and then, in a second step, absorbed by the spin. The second protocol, instead, will study the transfer from the external field through the cavity to the spins in a single step, where we consider exponential and Gaussian pulse shapes as well as two analytical and numerical approaches to pulse-shape optimisation. 

\section{Two-step ensemble absorption via intermediary cavity excitation}
\label{IV}

%%%%%%%%%%%%%%%%%%%%%%%%%%%%%%%%%%%%
\begin{figure*}[t!]
 \includegraphics[width=.39\textwidth]{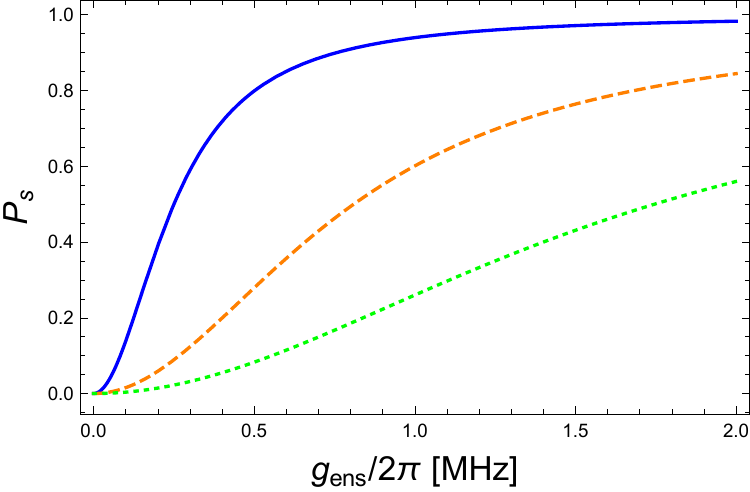}
 \hspace{1cm}
 \includegraphics[width=.39\textwidth]{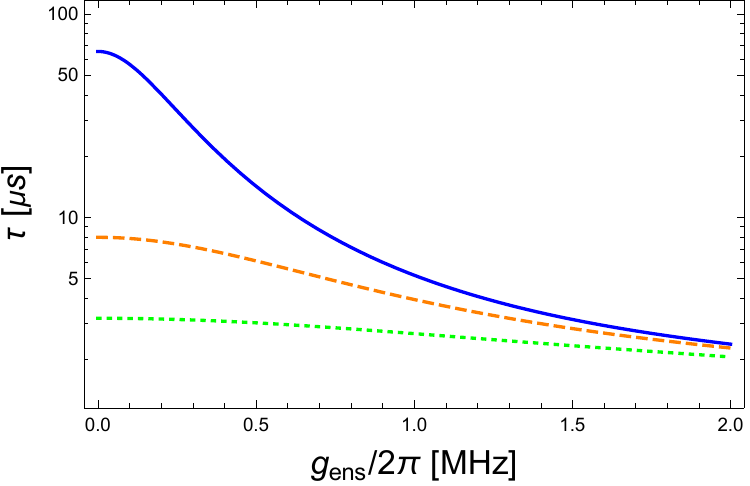}
\caption{Two-step photon-absorption protocol. Left panel: The probability $P_s$ as a function of the ensemble-coupling constant $g_{\text{ens}}$; see Eq.~\eqref{eq:twostepexcitation}. 
Right panel: Semilogarithmic plot of the 
required time $\tau$ of the two-step protocol to reach the maximum probability, where $\kappa_{\text{max}}/2 \pi = 1$ MHz; see Eq.~\eqref{eq:kappa}. In the numerical evaluation of the 
probability, we have considered $e^{-5} \approx 0$. The width of the inhomogeneous broadening is set to 
$w/2\pi = 10$ MHz. The curves belong to different values of $\kappa_{\text{min}}$: $\kappa_{\text{min}}/2 \pi = 25$ kHz (solid), $\kappa_{\text{min}}/2 \pi = 250$ kHz (dashed), and 
$\kappa_{\text{min}}/2 \pi = 1$ MHz (dotted).}
   \label{fig:twostep}
\end{figure*}
%%%%%%%%%%%%%%%%%%%%%%%%%%%%%%%%%%%%

The basic idea of the two-step protocol is to split the dynamical evolution into two parts: in the first step the tunable parameters are set to values that guarantee that the interaction between the spin ensemble and the single-mode 
cavity field is suppressed and the photon is stored in the cavity field; in the last step the interaction between the spin ensemble and the cavity field is turned on and now the photon in the cavity can be 
absorbed by the spins. 

Provided that a fast modulation of $\omega_c$ and $\kappa$ compared to the time-evolution of system is possible, we define
the two-step protocol as
\begin{equation}
 \Delta_{cs}(t)=\begin{cases} \Delta_{cs}, & t \in [0,t_0] \\ 0, & t \in (t_0,\infty) \end{cases} \label{eq:dcs}
\end{equation}
and
\begin{equation}
 \kappa(t)=\begin{cases} \kappa_{\text{max}}, & t \in [0,t_0] \\ \kappa_{\text{min}}, & t \in (t_0,\infty) \end{cases}. \label{eq:kappa}
\end{equation}
In other words, we assume that the values of $\Delta_{cs}(t)$ and $\kappa(t)$ are constant during the two steps of the protocol and their values can be switched instantaneously between the first and the second step.

\subsection{Cavity excitation}

To switch off interaction between the spins and the cavity field we require a condition on $\Delta_{cs}$. We obtain the general solution to Eq.~\eqref{eq:cfinal} with the help of the Laplace transform and its inverse. Furthermore, we use the fact that the Laplace transform of a convolution is simply the product of the individual transforms \cite{Davies}. The solution reads
\begin{eqnarray}
 \Psi_c(t)&=&\int^t_0\, dt' e^{-(2i \Delta_{cs}+ \kappa + w) (t-t')/4} \left \{ \cosh \left[\varpi (t-t')/4\right] \right. \nonumber \\
 &-& \left. \frac{2 i \Delta_{cs} + \kappa -w}{\varpi} \sinh \left[\varpi (t-t')/4\right] \right\} \alpha f(t') \label{eq:csol}
\end{eqnarray}
where
\begin{equation}
 \varpi=\sqrt{(2i \Delta_{cs}+ \kappa -w)^2-16 g^2_{\text{ens}}}. \label{eq:C}
\end{equation}
Moreover, if
\begin{equation}
 \left| \frac{4g_{\text{ens}}}{2i \Delta_{cs}+ \kappa -w} \right| \ll 1 \label{eq:twostepcond}
\end{equation}
and making use of the general condition in \eqref{eq:imcond},
then $\varpi\approx 2i \Delta_{cs}+ \kappa -w$ and \eqref{eq:csol} simplifies to
\begin{equation}
 \Psi_c(t)=\alpha \int^t_0\, dt' e^{-(i \Delta_{cs}+ \kappa/2 ) (t-t')}f(t'), \nonumber
\end{equation}
which is the solution to
\begin{equation}
 \dot{\Psi}_c=-i \Delta_{cs} \Psi_c+\alpha f(t) - \frac{\kappa}{2}\Psi_c, \label{eq:1step}
\end{equation}
with switched-off spin interactions.

Now, provided that $\Delta_{cs}$ and $\kappa$ fulfil condition \eqref{eq:twostepcond} for the experimentally fixed values of $g_{\text{ens}}$ and $w$, we consider the first step of the protocol 
for $t \in [0,t_0]$. 
The unnormalised pulse
\begin{equation}
 g(t) =\left(2i\Delta_{cs}+\kappa\right) H(t_0-t) e^{\left(i\Delta_{cs}+\kappa/2\right)(t-t_0)}, \label{eq:f2stepp}
\end{equation}
with the Heaviside function $H(x)$ would lead to an ideal cavity excitation 
\begin{equation}
  \Psi_c (t_0)=1, \quad \text{and} \quad  \sum_{j \in L} |\Psi^{(j)}_E(t_0)|^2=0, 
\end{equation}
and provides the unique solution 
\begin{equation}
  \Psi_c(t)=e^{-\left(i\Delta_{cs}+\kappa/2\right)|t-t_0|}
\end{equation}
with $\kappa t_0 /2 \gg 1$, i.e., $e^{-\kappa t_0 /2}\approx 0$. This solution is valid for time $t \in [0, \infty)$, but we restrict it to the interval $[0,t_0]$ and this is nothing else than the application of the well-known time-reversal approach, which is employed to perfectly excite an atomic state \cite{Cirac, Leuchs2009, Scarani2011, Korotkov, Kurtsiefer2013, Wenner, Leuchs2014}.
Using the constraint in Eq. \eqref{eq:constraint}, the pulse of the external drive reads
\begin{equation}
    f_e(t) = \alpha \underbrace{\frac{\kappa/2}{\sqrt{\Delta^2_{cs}+\kappa^2/4}}}_{=r} g(t) \label{eq:f2step}
\end{equation}
which leads to the physical excitation 
\begin{equation}
  \Psi_c (t_0)=\alpha r.
\end{equation}

Furthermore, for a given density of states, $\rho_E(\mathbf{k})$,
Eq.~\eqref{eq:initial} together with Eq.~\eqref{eq:f2step} determines the initial probability amplitudes $\Psi_E(\mathbf{k}, 0)$ of the external field. If the orthonormal mode functions of the external field are 
known then it is possible to obtain the characteristic shape of the incoming one-photon wave packet. If we consider that the value of $\kappa$ can be varied between 
$\kappa_{\text{min}}$ and $\kappa_{\text{max}}$, then to have a fast evolution of the first step $\kappa$ ought to be equal to $\kappa_{\text{max}}$ for $t \in [0,t_0]$. 

\subsection{Absorption into the ensemble}

During the second step of the protocol we turn on the interaction between the single-mode cavity field and the spin ensemble, i.e., $\Delta_{cs}=0$ for $t \in (t_0, \infty)$. As we have $\Psi_c (t_0)=\alpha r$, we require
$\kappa=\kappa_{\text{min}}$ or in other words $\kappa_e$ is small as possible, i.e., the escape of the photon from the cavity is reduced.

In the second step of the protocol, we have $\sum_{j \in L} |\Psi^{(j)}_E(t_0)|^2=0$ and Eqs. \eqref{eq:s}, \eqref{eq:c}, and \eqref{eq:E} yield
\begin{equation}
 \dot{\Psi}_c=-g^2_{\text{ens}} \int^t_{t_0}  e^{-w (t-t')/2} \Psi_c(t')\,dt' - \frac{\kappa_{\text{min}}}{2}\Psi_c.
\end{equation}
The solution for $t>t_0$ is
\begin{eqnarray}
 \Psi_c(t)&=& \alpha re^{-( \kappa_{\text{min}} + w) (t-t_0)/4} \left \{ \cosh \left[\varpi' (t-t_0)/4\right]  \right. \nonumber \\
 &-&\left. \frac{ \kappa_{\text{min}} -w}{\varpi'} \sinh \left[\varpi' (t-t_0)/4\right] \right\}, \label{eq:c2step}
\end{eqnarray}
where
\begin{equation}
 \varpi'=\sqrt{-16 g^2_{\text{ens}}+(\kappa_{\text{min}}-w)^2}.
\end{equation}
It is worth noting, for general parameters, $\varpi'$ can also be an imaginary number, which yields oscillations in Eq.~\eqref{eq:c2step}, i.e, both hyperbolic functions become trigonometric ones. However, based on the condition in \eqref{eq:imcond}, the subsequent analysis is done only for real values of $\varpi'$. Now, the solution for $\Psi^{(i)}_s$, similarly to Eq.~\eqref{eq:psis-integrodifferential}, reads 
\begin{equation}
 \bar{\Psi}^{(i)}_s(t)=-i g_i \int^t_{t_0} e^{i (\Delta_N -\Delta_i) (t-t')} \Psi_c(t')\,dt',
\end{equation}
and if we consider a long enough interaction time $t$ such that also the real part of the slowest decaying exponential vanishes, i.e., $\operatorname{Re}[\kappa_{\text{min}}+w-\varpi'] (t-t_0)/4 \gg 1$, then
\begin{eqnarray}
 &&\bar{\Psi}^{(i)}_s(t)= \\
 &&\frac{\alpha rg_i e^{i (\Delta_N -\Delta_i) (t-t_0)} (4 \Delta_i -4 \Delta_N+2 i w)}{(2\Delta_i -2\Delta_N +iw) (2\Delta_i -2\Delta_N+ i \kappa_{\text{min}})-4 g^2_{\text{ens}}}. \nonumber
\end{eqnarray}
The excitation probability of the spin ensemble reads
\begin{eqnarray}
 &&P_{s,\alpha r}=\alpha^2 r^2P_s=\sum^N_{i=1} |\Psi^{(i)}_s(t) |^2= \nonumber \\
 &&=\int^\infty_{-\infty}\, d\Delta \int^\infty_{-\infty}\, d g
 p_1(\Delta)  p_2(g) |\Psi_s(\Delta,g,t) |^2 \nonumber \\
 &&= \alpha^2 r^2\frac{4 g^2_{\text{ens}} w}{(\kappa_{\text{min}} +w)(4 g^2_{\text{ens}} + \kappa_{\text{min}} w)}, \label{eq:twostepexcitation}
\end{eqnarray}
where
\begin{equation}
\Psi_s(\Delta,g,t)=\frac{\alpha rg e^{-i \Delta (t-t_0)} (4 \Delta +2 i w)}{(2\Delta  +iw) (2\Delta + i \kappa_{\text{min}})-4 g^2_{\text{ens}}}. \nonumber
\end{equation}

The formula in \eqref{eq:twostepexcitation} gives the maximum, ideal amount of excitation probability $P_s$. Furthermore, the diminishing effects of the detuning $\Delta_{cs}$ and the internal cavity loss $\kappa_i$ are included through $r$ and $\alpha$ respectively\footnote{Note that $r(\Delta_{cs}=0)=1$, $ \alpha(\kappa_i=0)=1$, and $r, \alpha \leq 1$.}, and culminate in the excitation probability $P_{s,\alpha r}$. The ideal excitation probability $P_s$ can also be expressed as a function of the cooperativity $C=4 g^2_{\text{ens}}/(\kappa_{\text{min}} w)$
\begin{equation}
P_s=\frac{C}{(1+C)(1+\kappa_{\text{min}}/w)}, \nonumber
\end{equation}
which in the useful limiting case $w \gg \kappa_{\text{min}}$ yields $P_s=C/(1+C)$. In this limit, we want $C$ as high as possible.
 
\begin{figure}[t!]
 \includegraphics[width=.49\textwidth]{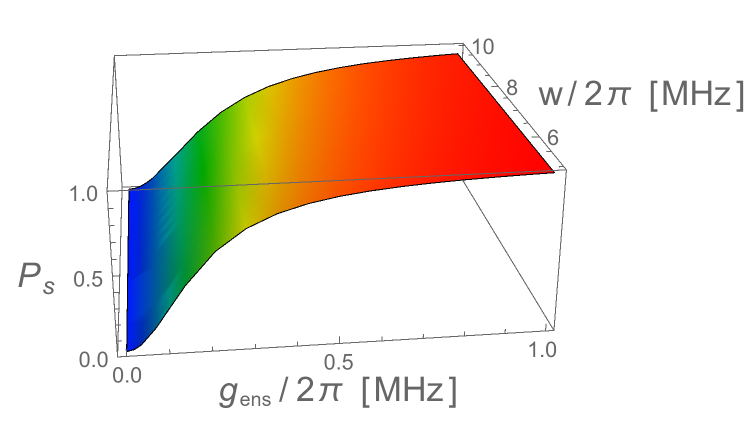}
\caption{Two-step protocol. The probability $P_s$ as a function of the ensemble-coupling constant $g_{\text{ens}}$ and the width of the inhomogeneous broadening $w$; 
see Eq.~\eqref{eq:twostepexcitation}. The decay rate of the cavity is set to $\kappa/2 \pi = 25$ kHz. }
   \label{fig:twostep3D}
\end{figure}

In Fig. \ref{fig:twostep} it is shown how the probability $P_s$  in \eqref{eq:twostepexcitation} changes for different values of $g_{\text{ens}}$ and $\kappa_{\text{min}}$. We have also 
plotted the required time to reach these values of the probability. There is a tradeoff between getting high probabilities and reaching them as fast as possible, see for example the solid and the dotted curves in Fig. \ref{fig:twostep}. We have argued earlier for 
a fast enough protocol to avoid population and phase decays of the spins and therefore the consistency of \eqref{eq:twostepexcitation} has to be always examined for given experimental values of the parameters. 
In general, larger ensemble-coupling constants yield higher probabilities, while $\kappa_{\text{min}}$ and $w$ have destructive effects on the catch of the photon by the spin ensemble. According to 
Fig. \ref{fig:twostep}, large values of $g_{\text{ens}}$ not only yield good absorptions of the photon but also a fast protocol. Fig. \ref{fig:twostep3D} shows the dependence of the  probability 
$P_s$ on the width of the inhomogeneous broadening $w$; as the value of $w$ is decreasing $P_s$ is slightly increased. Inhomogeneous broadening of the spin ensemble is always an obstacle from the point of 
view of controllability, but here the two-step protocol has a reduced impact on the excitation storage.   

%%%%%%%%%%%%%%%%%%%%%%%%%%%%%%%%%%%%
%%%%%%%%%%%%%%%%%%%%%%%%%%%%%%%%%%%%
\subsection{Gaussian and other distributions of detunings}\label{sec:other_distributions}

\begin{figure}[h!]
 \includegraphics[width=.39\textwidth]{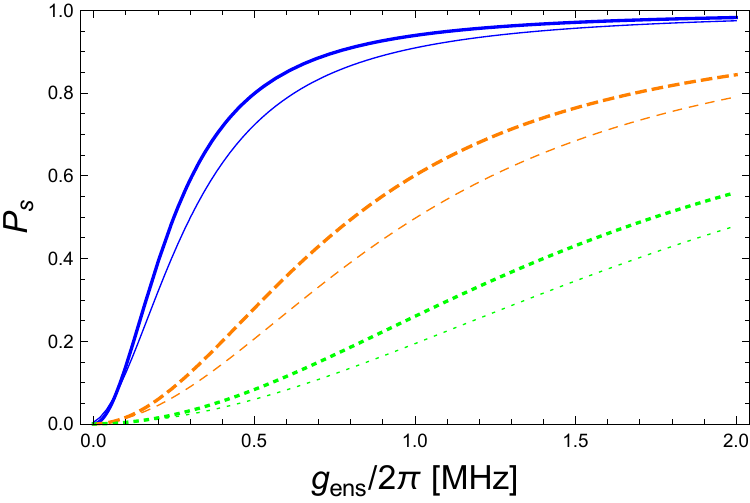}
\caption{Two-step protocol with Lorentzian (thick) and Gaussian (thin) broadened spin ensembles. The probability $P_s$ as a function of the ensemble-coupling constant $g_{\text{ens}}$. All parameters and settings correspond
to those of Fig. \ref{fig:twostep}.}
   \label{fig:twostepG}
\end{figure}

To demonstrate how the results from above can be generalised to arbitrary distributions of the inhomogeneous broadening of the spins, we first consider a Gaussian broadened spin ensemble
 \begin{equation}
p_1(\Delta)=\frac{1}{\sqrt{2 \pi} w } e^{-\Delta^2/
(2 w^2)}. \label{eq:Gaussd}
\end{equation}
Based on the method described in Appendix \ref{AppendixA}, we approximate this probability distribution with the sum of eight Lorentzian-shaped functions. These functions enable us to find numerically the poles required for the analytical evolution of the inverse Laplace transform. The results are presented in Fig.~\ref{fig:twostepG}. The probability $P_s$ of the Gaussian broadened spin ensemble is slightly smaller, but, in general, both distributions deliver the same features of photon absorption.

In typical experiments, the distribution may have somewhat arbitrary broadening, including potentially multimodal distributions. For situations that differ significantly from Lorentzians or Gaussians, the approximation method discussed above can still give closed-form solutions to the temporal dynamics.

%%%%%%%%%%%%%%%%%%%%%%%%%%%%%%%%%%%%
%%%%%%%%%%%%%%%%%%%%%%%%%%%%%%%%%%%%

It is worth noting that we have studied the properties of the probability $P_s$, but the excitation probability in Eq. \eqref{eq:twostepexcitation} is obtained by multiplying $P_s$ with $\alpha^2 r^2$. This number is smaller than one, because $\Delta_{cs}>0$, and usually $\kappa_i>\kappa_e$ \cite{Wen2022}.

\section{One step, direct photon absorption}
\label{V}

In this scenario, we investigate the possibility of direct absorption of the photon by the spin ensemble, i.e., the interaction between the single-mode cavity field and the spins is never turned off. Thus,  
the central spin frequency is equal for all times to the frequency of the single-mode field, which means that $\Delta_{cs}=0$.
We first consider the one-step protocol as a special case of the two-step protocol. We then consider analytically derivable pulse shapes and finally at Gaussian pulses and optimal-control pulses.
The optimised pulsed shapes will allow us to find also respective optimal values for the decay rate $\kappa$. Based on the model of Sec. \ref{II} the main coupled differential equations are
\begin{eqnarray}
  \dot{\Psi}^{(i)}_s &=&i (\Delta_N -\Delta_i)  \Psi^{(i)}_s -i g_i \Psi_c, \nonumber \\   
  \dot{\Psi}_c&=&-g^2_{\text{ens}} \int^t_0  e^{-w (t-t')/2} \Psi_c(t')\,dt'+\alpha f(t) - \frac{\kappa}{2}\Psi_c. \nonumber 
\end{eqnarray}
If $\kappa$ is considered to be constant, then the solution for $\Psi^{(i)}_s$ reads
\begin{eqnarray}
 &&\Psi^{(i)}_s(t)= \alpha g_i \int^t_0\, dt' f(t') \left \{ A e^{-\frac{\kappa + w}{4} (t-t')} \cosh \left[\frac{\varpi'}{4} (t-t')\right] \right. \nonumber \\
 &&+ \left. B e^{-\frac{\kappa + w}{4} (t-t')} \sinh \left[\frac{\varpi'}{4} (t-t')\right]-A e^{i (\Delta_N -\Delta_i) (t-t')}  \right\}, \nonumber \\
 \label{eq:onestepspin}
\end{eqnarray}
where
\begin{eqnarray}
 A&=& \frac{ 4 \Delta_N -4 \Delta_i-2 i w}{(2\Delta_i -2\Delta_N +iw) (2\Delta_i -2\Delta_N+ i \kappa)-4 g^2_{\text{ens}} }, \nonumber \\
 B&=& \frac{1}{\varpi'}\frac{ (4 \Delta_N -4 \Delta_i-2 i w) (w-\kappa)+16i g^2_{\text{ens}}}{(2\Delta_i -2\Delta_N +iw) (2\Delta_i -2\Delta_N+ i \kappa)-4 g^2_{\text{ens}} }. \nonumber
\end{eqnarray}

\subsection{Exponential shape pulse}

In this subsection, we demonstrate that the one-step and two-step protocols are markedly different. To understand the situation better we consider 
\begin{equation}
 f(t)= \kappa H(t_0-t) e^{\kappa(t-t_0)/2}, \label{eq:f1step}
\end{equation}
which is the ideal solution for the two-step protocol with $\Delta_{cs}=0$, see Eq.~\eqref{eq:f2step}. This choice guarantees that the single-mode cavity field is  excited at $t=t_0$ with probability $\alpha^2$. Therefore, we 
have evaluated in this one-step protocol the excitation probability of the cavity field at $t=t_0$ and a lengthy calculation involving the integration over the detunings of the spin ensemble with the Lorentzian weight 
$p_1(\Delta)$ yields
\begin{equation}
 |\Psi_c(t_0)|^2=\alpha^2\frac{\kappa^2 (\kappa +w)^2}{\left[2 g^2_{\text{ens}} + \kappa (\kappa + w) \right]^2}. \label{eq:1stepcavt0}
\end{equation}
Similarly, the excitation is in the spin ensemble with probability:
\begin{equation}
 \alpha^2P_s(t_0)=\sum^N_{i=1} |\Psi^{(i)}_s(t_0) |^2=\alpha^2\frac{4 g^2_{\text{ens}} \kappa (\kappa +w)}{\left[2 g^2_{\text{ens}} + \kappa (\kappa + w) \right]^2}. \label{eq:1stepspint0}
\end{equation}
Both formulas are valid under the assumption $\kappa t_0 /2 \gg 1$ or $e^{-\kappa t_0 /2}\approx 0$. It is worth noting that the maximum value taken by $ P_s(t_0)$ is $0.5$. In general, when the external drive in Eq. \ref{eq:f1step} ends at $t=t_0$, the sum of the probabilities in Eqs. \ref{eq:1stepcavt0} and \ref{eq:1stepspint0} can not be $\alpha^2$, unless $g_{\text{ens}}=0$, i.e., the spins do not interact with the field, which resembles the first phase of the two-step protocol.

If we consider a long enough interaction time $t$ such that $\operatorname{Re}[\kappa+w-\varpi'] (t-t_0)/4 \gg 1$, then $\Psi_c(t)\approx 0$ and the excitation probability of the spin ensemble is 
\begin{eqnarray}
 &&P_{s,\alpha}=\alpha^2P_s=\sum^N_{i=1} |\Psi^{(i)}_s(t) |^2= \int^\infty_{-\infty} d\Delta\, p_1(\Delta) \label{eq:1stepfinalspin} \\
 &&\times \frac{16 \alpha^2 g^2_{\text{ens}} \kappa^2 (4 \Delta^2 +w^2)}{(4 \Delta^2 +\kappa^2) \left[4 \Delta^2 (\kappa+w)^2 + (4 g^2_{\text{ens}}+\kappa w-4\Delta^2)^2 \right]}. \nonumber
\end{eqnarray}
The above integration can be analytically done, however it yields cumbersome formulas, which are not worth being presented. 
%%%%%%%%%%%%%%%%%%%%%%%%%%%%%%%%%%%%
\begin{figure}[t!]
 \includegraphics[width=.39\textwidth]{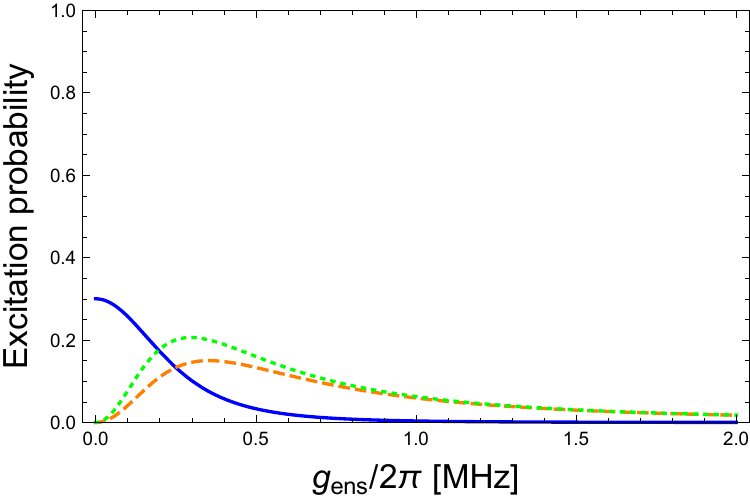}
\caption{One-step protocol with $f(t)$ in Eq.~\eqref{eq:f1step}.  Excitation probabilities of the spin ensemble and the single-mode cavity field as a function of the ensemble-coupling constant $g_{\text{ens}}$. 
The width of the inhomogeneous broadening is set to 
$w/2\pi = 10$ MHz, $\kappa/2 \pi = 25$ kHz, and $\alpha^2=0.3$ \cite{Bertet2020} The curves show different probabilities: the cavity field is excited at $t=t_0$, see Eq.~\eqref{eq:1stepcavt0} (solid); 
the spin ensemble absorbed the excitation at $t=t_0$, see Eq.~\eqref{eq:1stepspint0} (dashed); the spin ensemble absorbed the excitation after a longer interaction time, when the cavity is empty,  
see Eq.~\eqref{eq:1stepfinalspin} (dotted).}
   \label{fig:onestep1}
\end{figure}
%%%%%%%%%%%%%%%%%%%%%%%%%%%%%%%%%%%%

\begin{figure}[t!]
 \begin{center}
 \includegraphics[width=.39\textwidth]{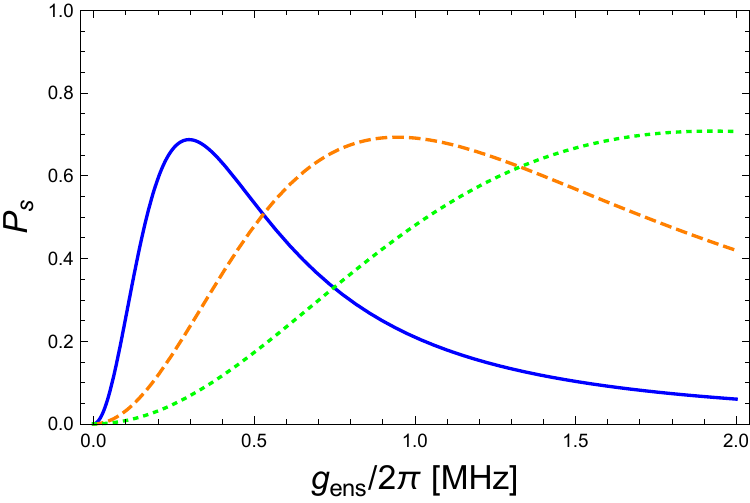}
 \includegraphics[width=.39\textwidth]{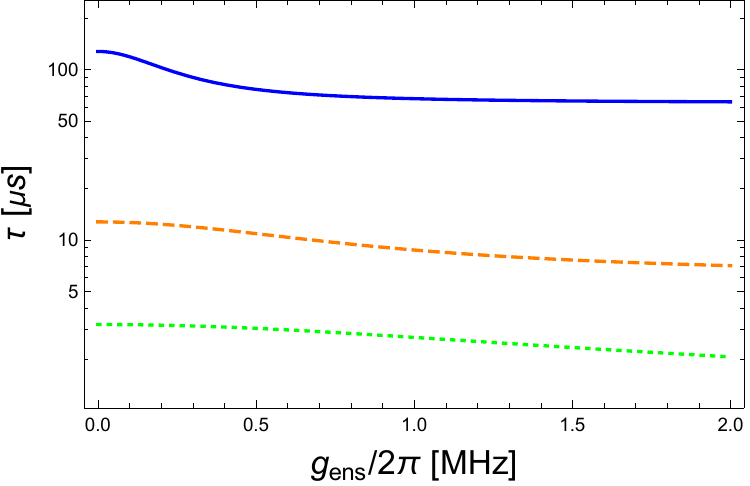}
\caption{One-step protocol with $f(t)$ in Eq.~\eqref{eq:f1step}. Top panel: The probability $P_s$ of Eq. \eqref{eq:1stepfinalspin} as a function of the 
ensemble-coupling constant $g_{\text{ens}}$. Bottom panel:  Semilogarithmic plot of the required time $\tau$ of the protocols to reach the maximum probability. All parameters and settings correspond to those of Fig. \ref{fig:twostep}, with as before the lines corresponding to $\kappa/2 \pi = 25$ kHz (solid), $\kappa/2 \pi = 250$ kHz (dashed), and 
$\kappa/2 \pi = 1$ MHz (dotted).}
   \label{fig:onevstwo}
 \end{center}
\end{figure}

\begin{figure}[t]
 \includegraphics[width=.49\textwidth]{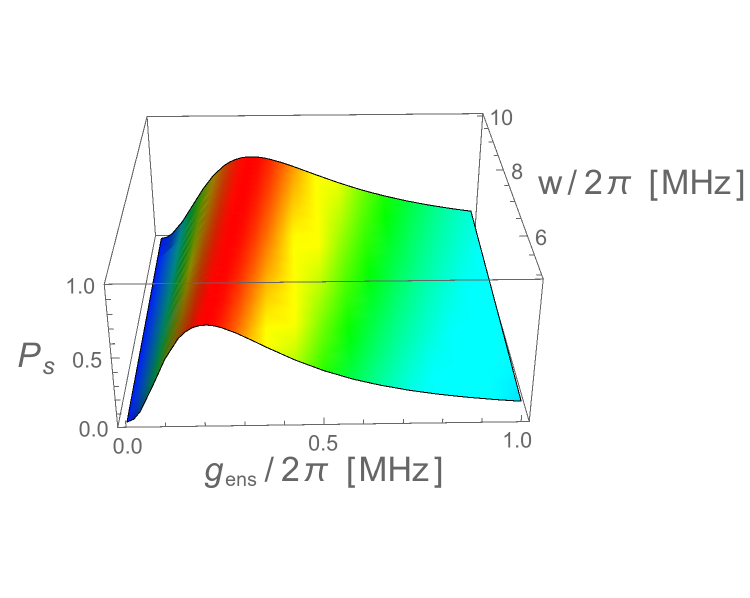}
\caption{One-step protocol with $f(t)$ in Eq.~\eqref{eq:f1step}. The probability $P_s$ as a function of the ensemble-coupling constant $g_{\text{ens}}$ and the width of the 
inhomogeneous broadening $w$; see Eq.~\eqref{eq:1stepfinalspin}. The decay rate of the cavity is set to $\kappa/2 \pi = 25$ kHz. }
   \label{fig:onestep3D1}
\end{figure}

The choice of $f(t)$ or indirectly the choice of the initial conditions of the external field results in the following effect: a larger ensemble-coupling constant $g_{\text{ens}}$ does not necessarily imply 
better probabilities $P_s$ of the spin ensemble. This is apparent from Fig. \ref{fig:onestep1}. When the cavity field becomes empty, the probability $P_s$ in \eqref{eq:1stepfinalspin} is always larger than the one in \eqref{eq:1stepspint0}, because the photon at $t=t_0$ leaks for later times into both the spin ensemble and the external field or gets lost inside the cavity. The single-mode of the cavity can be well excited at $g_{\text{ens}}=0$ and $P_s$ reaches its maximum after the cavity is empty and only for a given value of $g_{\text{ens}}$. This value is increased with 
the increase of $\kappa$. This repulsive character of the joint system of the cavity field and the spin ensemble is very 
different from the results obtained for the two-step protocol; see Fig. \ref{fig:twostep}. For comparison, we show the results for the exponential pulse for the one-step protocol in Fig. \ref{fig:onevstwo}. The one-step protocol performs better for low values of
$g_{\text{ens}}$, but it requires longer times to reach better probabilities of the spin ensemble. For shorter protocols, a larger $\kappa$ is needed, but then the maximum of the probability is shifted 
towards larger values of $g_{\text{ens}}$. Whether a faster protocol or better absorption of the incoming photon is preferred, depends on the experimental setup and the planned further control of the 
spin ensemble, e.g., the application of $\pi$ pulses to refocus the spin dephasing. A reduced value of the inhomogeneous broadening $w$ shifts the maximum of $P_s$
towards smaller values of $g_{\text{ens}}$; see Fig. \ref{fig:onestep3D1}. This is not surprising, because larger $w$ mean more far-detuned spin transitions, which limit the storage of the photon in the spin 
ensemble. 

%%%%%%%%%%%%%%%%%%%%%%%%%%%%%%%%%%%%
%%%%%%%%%%%%%%%%%%%%%%%%%%%%%%%%%%%%

%%%%%%%%%%%%%%%%%%%%%%%%%%%%%%%%%%%%
%%%%%%%%%%%%%%%%%%%%%%%%%%%%%%%%%%%%

\subsection{Analytically derivable pulse shape}

As we have demonstrated, reusing the exponential pulse $f(t)$ derived for the two-step protocol is not a practical choice to store the excitation in the spin ensemble in one step. Therefore, in the subsequent discussion, we investigate further candidates, which lead to more 
optimal storage. First, we rewrite Eq.~\eqref{eq:onestepspin} as
\begin{equation}
 \Psi^{(i)}_s(t)= \alpha g_i \int^t_0\, dt' f(t') h_i(t',t), \nonumber
\end{equation}
where the integral kernel $h_i(t',t)$ is implicitly defined via Eq.~\eqref{eq:onestepspin}.
The excitation probability of the spin ensemble at time $T$ reads
\begin{eqnarray}
 &&P_{s,\alpha}=\alpha^2P_s=\sum^N_{i=1} |\Psi^{(i)}_s(T) |^2 \label{eq:inneruse}\\
 &&= \alpha^2g^2_{\text{ens}} \int^\infty_{-\infty} d\Delta\, p_1(\Delta)  \left | \int^T_0\, dt' f(t') h(t',T, \Delta) \right|^2, \nonumber
\end{eqnarray}
with
\begin{eqnarray}
 &&h(t',T, \Delta)=A(\Delta) e^{-\frac{\kappa + w}{4} (T-t')} \cosh \left[\frac{\varpi'}{4} (T-t')\right] \nonumber \\
 &&+ B(\Delta) e^{-\frac{\kappa + w}{4} (T-t')} \sinh \left[\frac{\varpi'}{4} (T-t')\right]-A(\Delta) e^{-i\Delta (T-t')} \nonumber
\end{eqnarray} 
and
\begin{eqnarray}
A(\Delta)&=&\frac{-4 \Delta-2 i w}{(2\Delta +iw) (2\Delta+ i \kappa)-4 g^2_{\text{ens}} }, \nonumber \\
B(\Delta)&=&\frac{1}{\varpi'}\frac{ (-4 \Delta-2 i w) (w-\kappa)+16i g^2_{\text{ens}}}{(2\Delta +iw) (2\Delta+ i \kappa)-4 g^2_{\text{ens}}}. \nonumber
\end{eqnarray}

%%%%%%%%%%%%%%%%%%%%%%%%%%%%%%%%%%%%
\begin{figure}[t!]
 \begin{center}
 \includegraphics[width=.39\textwidth]{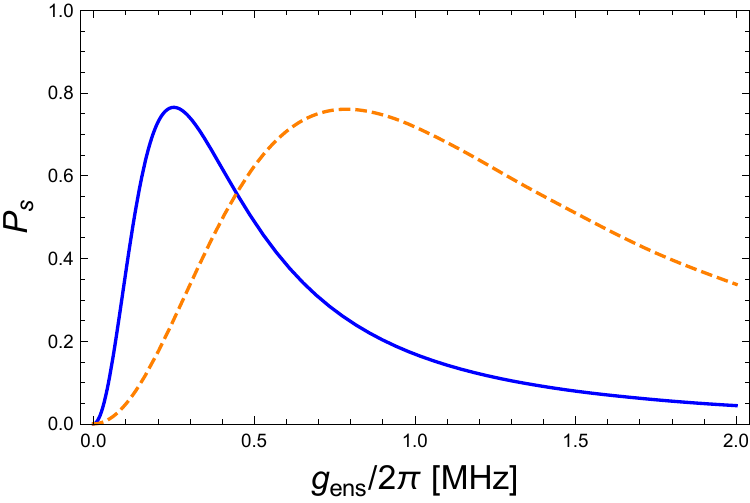}
 \includegraphics[width=.39\textwidth]{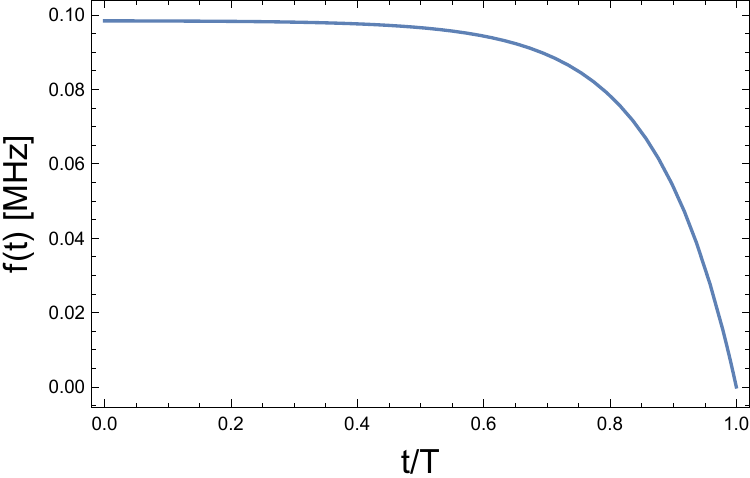}
\caption{Optimised one-step protocol with $f(t)$ in Eq.~\eqref{eq:f1stepmoreopt}. Top panel: The probability $P_s$ of Eq. \eqref{eq:inneruse} as a function of the 
ensemble-coupling constant $g_{\text{ens}}$. The curves belong to different values of $\kappa$: $\kappa/2 \pi = 25$ kHz (solid) and $\kappa/2 \pi = 250$ kHz (dashed). 
Bottom panel: The optimal $f(t)$ as a function of time, where $T=20 \mu$s and $\kappa/2 \pi = 25$ kHz. The width of the inhomogeneous broadening is set to 
$w/2\pi = 10$ MHz.}
   \label{fig:onemoreopt}
 \end{center}
\end{figure}
%%%%%%%%%%%%%%%%%%%%%%%%%%%%%%%%%%%%

We observe that $h(t',T, \Delta) \in L^2\left([0,T]\right)$, i.e. $h$ is a square-integrable function on the interval $[0, T]$. The function space $L^2\left([0,T]\right)$ is a Hilbert space with 
the inner product of the physicist convention \cite{Yosida}
\begin{equation}
 \braket{f, h}=  \int^T_0\, dt' \bar{f} (t') h(t'), \nonumber
\end{equation}
where $\bar{z}$ denotes the complex conjugate of a complex number $z$. In the case of this particular inner product, the Cauchy-Bunyakovsky-Schwarz inequality reads
\begin{equation}
 \left |\int^T_0\, dt' \bar{f}(t') h(t') \right|^2 \leqslant \int^T_0\, dt' |f(t')|^2 \int^T_0\, dt' |h(t')|^2, \nonumber \\
\end{equation}
where the equality occurs if and only if one of $h(t')$, $f(t')$ is a scalar multiple of the other. However, in Eq.~\eqref{eq:inneruse} we have a $\Delta$-dependent family of inner products of the form
$\langle f , h(\Delta) \rangle$ and their squared absolute values are integrated over all $\Delta$ with the weight function $p_1(\Delta)$. Thus, the Cauchy-Bunyakovsky-Schwarz inequality can be applied for each
value of $\Delta$, which results in a $\Delta$-dependent $f(t')$, i.e., multiple optimal solutions and they depend on the detunings of the individual spins. Instead, we consider only one optimal solution at the 
maximum $\Delta =0$ of the weight function $p_1(\Delta)$ in \eqref{eq:broaden} and arrive at
\begin{eqnarray}
 &&f(t)= \lambda \left\{ a-a e^{-\frac{\kappa + w}{4} (T-t)} \cosh \left[\frac{\varpi'}{4} (T-t)\right] \right. \nonumber \\
 &&+ \left. b e^{-\frac{\kappa + w}{4} (T-t)} \sinh \left[\frac{\varpi'}{4} (T-t)\right]   \right\}, 
 \label{eq:f1stepmoreopt}
 \end{eqnarray}
where by absorbing the imaginary unit into $\lambda$ we get
\begin{eqnarray}
 a&=& \frac{2 w}{w\kappa+4 g^2_{\text{ens}} }, \nonumber \\
 b&=& \frac{1}{\varpi'}\frac{2 w (w-\kappa)-16 g^2_{\text{ens}}}{w\kappa+4 g^2_{\text{ens}}  }. \nonumber
\end{eqnarray}
The parameter $\lambda$ is found from the normalisation condition, which reads
\begin{equation}
 \int^T_0\, dt |f(t)|^2=\kappa. \label{eq:normalization}
\end{equation}
We consider again a long enough interaction time $T$ such that $\operatorname{Re}[\kappa+w-\varpi'] T/4 \gg 1$. In Fig. \ref{fig:onemoreopt}, we see the same behaviour of the probabilities $P_s$, which we 
have observed in Fig. \ref{fig:onestep1}, i.e., there is a maximum only for a given value of the ensemble-coupling constant $g_{\text{ens}}$. However, the curves 
in Fig. \ref{fig:onevstwo} reach a maximum value of $0.7$, whereas now the $P_s$ has a maximum larger than $0.75$. For the case of $\kappa/2 \pi = 25$ kHz, the optimal $f(t)$ is also displayed in Fig. \ref{fig:onemoreopt}. The required times to reach these excitation probabilities are the same as the ones 
depicted in Fig. \ref{fig:onevstwo}. The role of the inhomogeneous broadening $w$ is the same as what we have observed for the previous $f(t)$ in Eq.~\eqref{eq:f1step}; see again Fig. \ref{fig:onestep3D1}.
%%%%%%%%%%%%%%%%%%%%%%%%%%%%%%%%%%%%
\begin{figure}[t!]
 \begin{center}
 \includegraphics[width=.39\textwidth]{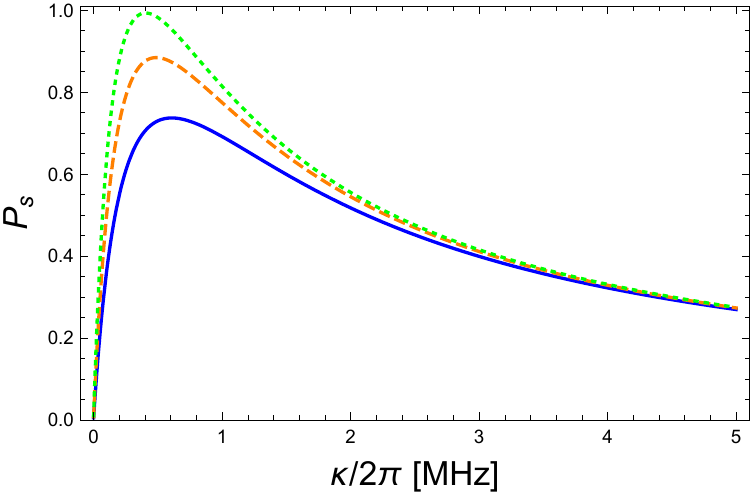}
 \includegraphics[width=.39\textwidth]{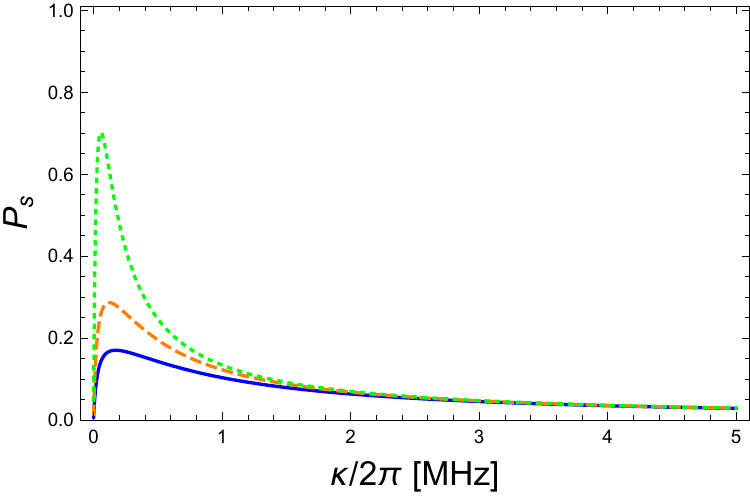}
 \includegraphics[width=.39\textwidth]{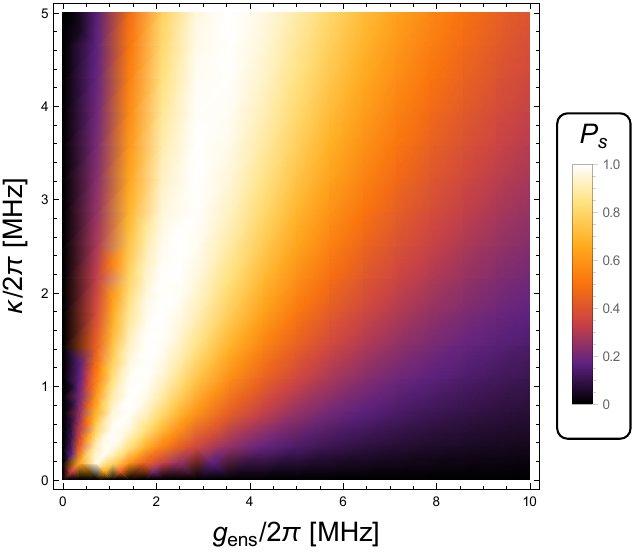}
\caption{The probability $P_s$ of Eq. \eqref{eq:1stepGauss} as a function of total cavity decay rate $\kappa$ with $g_{\text{ens}}/2 \pi = 1$ MHz (top panel) and $g_{\text{ens}}/2 \pi = 0.3$ MHz (middle panel). We set $\kappa_g$ in Eq.~\eqref{eq:f1stepGauss} as: $\kappa_g/2 \pi = 500$ kHz (solid); $\kappa_g/2 \pi = 250$ kHz (dashed);  and $\kappa_g/2 \pi = 50$ kHz (dotted). Bottom panel: Density plot of $P_s$ as a function of $\kappa$ and $g_{\text{ens}}$ for $\kappa_g/2 \pi = 50$ kHz.
The width of the inhomogeneous broadening is set for all figures to $w/2\pi = 10$ MHz.}
   \label{fig:Gauss}
 \end{center}
\end{figure}
%%%%%%%%%%%%%%%%%%%%%%%%%%%%%%%%%%%%
\subsection{Gaussian pulse shape}

\begin{figure*}[t!]
  \centering
  %two subfigures
  \begin{subfigure}[t]{0.45\textwidth}
    \includegraphics[width=8cm]{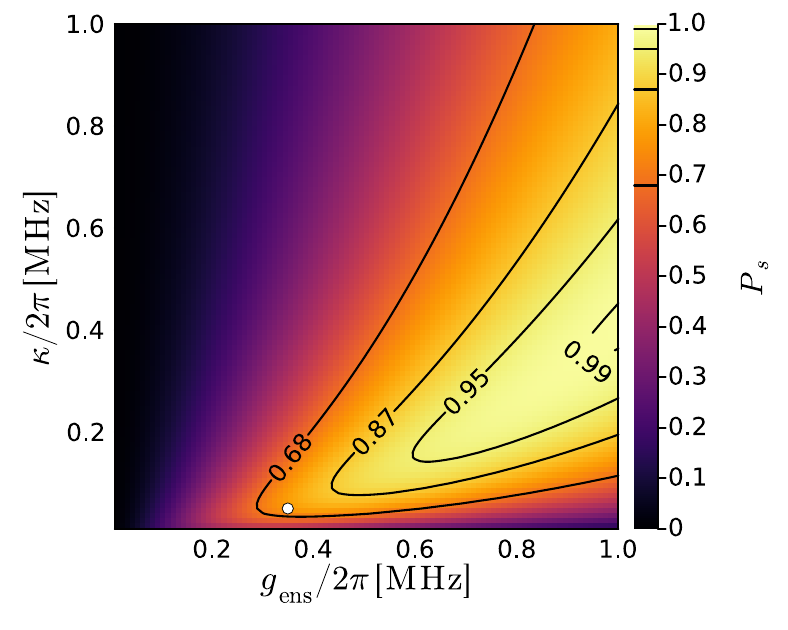}
    \caption{Gaussian pulse shapes}
    \label{fig:P_s_of_kappa_g_contour_zsolt}
  \end{subfigure}
  \hspace{1cm}
  \begin{subfigure}[t]{0.45\textwidth}
    \includegraphics[width=8cm]{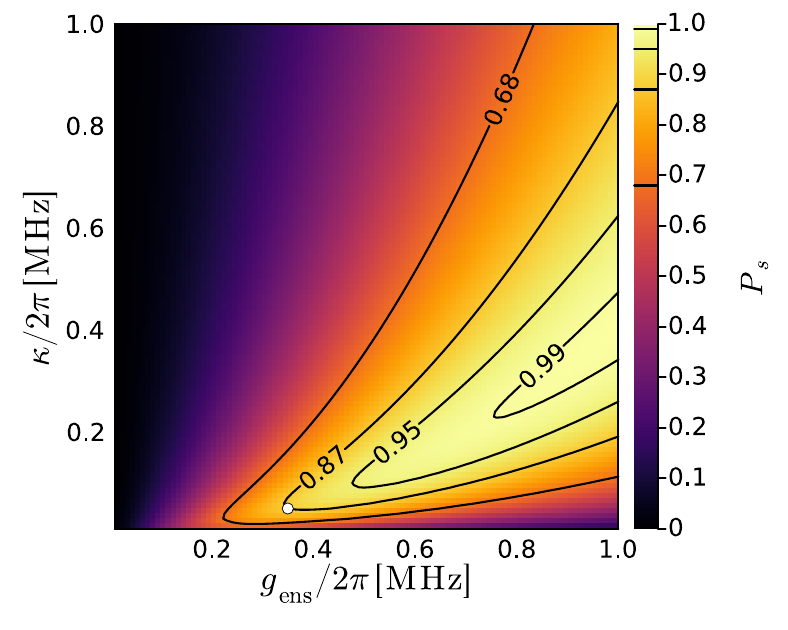}
    \caption{Optimal pulse shapes}
    \label{fig:P_s_of_kappa_g_contour_optim}
  \end{subfigure}
  \caption{The probability $P_s$ of Eq. \eqref{eq:inneruse} as a function of the total cavity decay rate $\kappa$ and ensemble-coupling constant $g_\text{ens}$ for Gaussian (a) and optimal (b) pulse shapes. We consider for both figures $\omega/2\pi= 10$ MHz and $T=20\mu$s. To highlight the higher values of $P_s$ for the optimal pulse shapes, we show in both plots a white dot at $g_\text{ens}/2\pi=350$ kHz and $\kappa/2\pi = 50$ kHz corresponding to a cooperativity of $C=0.98$ and observe an increase of $0.19$ in the value of $P_s$.
    \label{fig:P_s_of_kappa_g_contour}}
\end{figure*}

Finally, we investigate a widely used case when $f(t)$ is a Gaussian function, which can benefit from favourable bandwidth considerations \cite{Bertet2020}:
\begin{equation}
f(t)=\frac{\sqrt{\kappa \kappa_g}}{\pi^{\nicefrac{1}{4}}}e^{-\kappa^2_g (t-t_0)^2/2}, \label{eq:f1stepGauss}
\end{equation}
where $1/\kappa_g$ is the standard deviation of the pulse. If $t_0 \gg 1/\kappa_g$, then it is immediate that $\int^\infty_0\, dt |f(t)|^2=\kappa$. We substitute \eqref{eq:f1stepGauss} into Eq.~\eqref{eq:onestepspin} with both conditions $T > t_0$ and 
\begin{equation}
\operatorname{Re}\left[\frac{\kappa+w\pm\varpi'}{4} (T-t_0)- \frac{\left(\kappa+w \pm\varpi'\right)^2}{32 \kappa^2_g} \right]\gg 1 \label{eq:Gausscond}
\end{equation} being fulfilled. The latter condition is necessary for the approximation, where all $\Delta$-independent exponential terms are considered to be zero.
Based on this, the excitation probability $P_{s,\alpha}=\alpha^2P_s$ of 
the spin ensemble is obtained as before by replacing the summation over the spins with an integral involving the joint distribution $p_1(\Delta)p_2(g)$. Then, we have
\begin{eqnarray}
&&P_s= \int^\infty_{-\infty} d\Delta\, p_1(\Delta) 
\frac{2\kappa \sqrt{\pi}}{\kappa_g} e^{-\frac{\Delta^2}{\kappa^2_g}}\label{eq:1stepGauss} \\
 &&\times \frac{4 g^2_{\text{ens}}  (4 \Delta^2 +w^2)}{ \left[4 \Delta^2 (\kappa+w)^2 + (4 g^2_{\text{ens}}+\kappa w-4\Delta^2)^2 \right]}. \nonumber
\end{eqnarray}

In the first step, we investigate the width of the Gaussian and find numerically that optimal scenarios occur when the bandwidth $\kappa_g$ is as small as possible, see Fig. \ref{fig:Gauss}. In this figure, it is also demonstrated that good excitations are achieved for different decay rates of the cavity as a function of the ensemble-coupling $g_{\text{ens}}$. The repulsive behaviour of $P_s$ as a function of $g_{\text{ens}}$ is similar to the previous cases of the one-step protocol. The duration of the process depends on the standard deviation of the pulse $\sigma_g=1/\kappa_g$ and the condition in \eqref{eq:Gausscond}. First, $t_0 > 3 \sigma_g$ and $T>6 \sigma_g$ have to be valid such that the integration over time covers almost the whole pulse. In the case of $\kappa_g/2\pi= 50$ kHz, $\kappa/2 \pi=0.5$ MHz, $g_{\text{ens}}/2\pi=1$MHz, and 
$w/2\pi = 10$ MHz, we obtain $T>161.8\mu$s provided that
$e^{-5} \approx 0$. Thus, optimal excitation  is possible at the expense of an increase in the duration of the protocol. This is much longer than the duration times obtained for the two-step protocol, see Fig. \ref{fig:twostep}, where this set of parameters yields $T>3.28\mu$s. This raises the question of how can one obtain the best excitation scenario for considerably shorter times, which will be discussed in the subsequent section devoted to numerical analysis. 

\subsection{Optimal control - numerically optimised pulse shapes}
\label{sec:optnum}

We now turn our attention to the numerical optimisation for pulse shapes $f(t)$, decay rate of the cavity $\kappa$, and pulse durations $T$ to overcome the limitations of the previous attempts. 
In order to optimise constrained pulse shapes $f(t)$ (see. Eq.~\eqref{eq:normalization}) with otherwise given parameters
\begin{equation}
    f_\text{opt} = \underset{f}{\operatorname{argmax}}\, P_s\left(f, T, w, g_\text{ens}, \kappa  \right),\label{eq:f_opt_prob}
\end{equation}
we expand $f(t)$ in a set of basis functions $f(t) = \sum_{j=0}^{2 N_b} c_j f_j(t)$, where
\begin{eqnarray}
\label{eq:basisf}
  f_j(t) = \begin{cases}
    \cos\left(\pi \frac{j t}{T} \right)       
    & \text{for } 0\leqslant j \leqslant N_b, \\
    \sin\left(\pi \frac{(j-N_b) t}{T} \right) & \text{for } N_b<j \leqslant 2N_b. \label{eq:basis_function_definition}
  \end{cases}
\end{eqnarray}
We have $N_b+1$ cosine terms, with $f_0(t)=1$ being a constant term and $N_b$ sine terms. In some experimental setups \cite{Wen2022}, $\kappa$ can be adjusted, so we extend our approach to also optimise $\kappa$
\begin{equation}
    \kappa_\text{opt} = \underset{\kappa}{\operatorname{argmax}}\, P_s(f_\text{opt}(\kappa), T, w, g_\text{ens}, \kappa).\label{eq:kappa_opt_prob}
\end{equation}
Using the subsequent optimisation of pulse shape $f(t)$ and $\kappa$, we can also determine the shortest pulse duration $T_\text{min}$ needed to achieve a target probability $P_{s,\text{tar}}$. This is accomplished by identifying the root of the equation
\begin{equation}
    \Delta P_{s, \text{tar}}(T) = P_{s, \text{tar}} - P_s(f_\text{opt}(\kappa), T, w, g_\text{ens}, \kappa_\text{opt}), \label{eq:T_min_prob}
\end{equation}
where $\Delta P_{s, \text{tar}}(T)$ represents the deviation from the target probability. The condition $\Delta P_{s, \text{tar}} (T_\text{min}) = 0$ is sufficient for finding the minimum duration since we investigate a regime where the absorption is increasing monotonically with $T$. \\

We combine analytical and numerical techniques to solve Eq.~\eqref{eq:inneruse} restated for a set of basis functions, which allows us to solve the optimisation problems in Eqs.~(\ref{eq:f_opt_prob} - \ref{eq:T_min_prob}). The technical details of both optimisation and numerical integration are explained in Appendix \ref{sec:optimization_apendix}. The numerical approach was implemented using the Julia language \cite{bezanson2017julia}, our code is available at\cite{Michael}.\\

% optimized_pulse
\begin{figure}[t!]
  \centering
  \includegraphics[width=8cm]{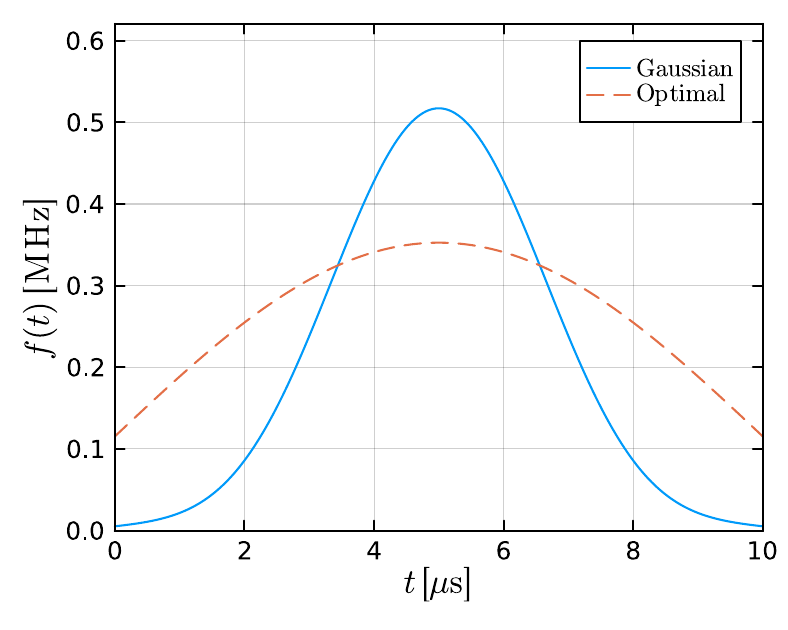}
  \caption{Gaussian (solid) and optimal (dashed) pulse shape for $g_\text{ens}/2\pi=0.5$ MHz, $T=10\mu$s, $w/2\pi=10$ MHz, and the Optimal $\kappa/2\pi\approx0.121$ MHz. For the Gaussian we set $\kappa_g = 2\pi/T$. We achieve a probability of $P_{s}=0.88$ for the Optimal and $0.78$ for the Gaussian pulse shapes.
  %($C=0.824$).
    \label{fig:optimal_pulse}}
\end{figure}

\begin{figure}
  \centering
  \includegraphics[width=8cm]{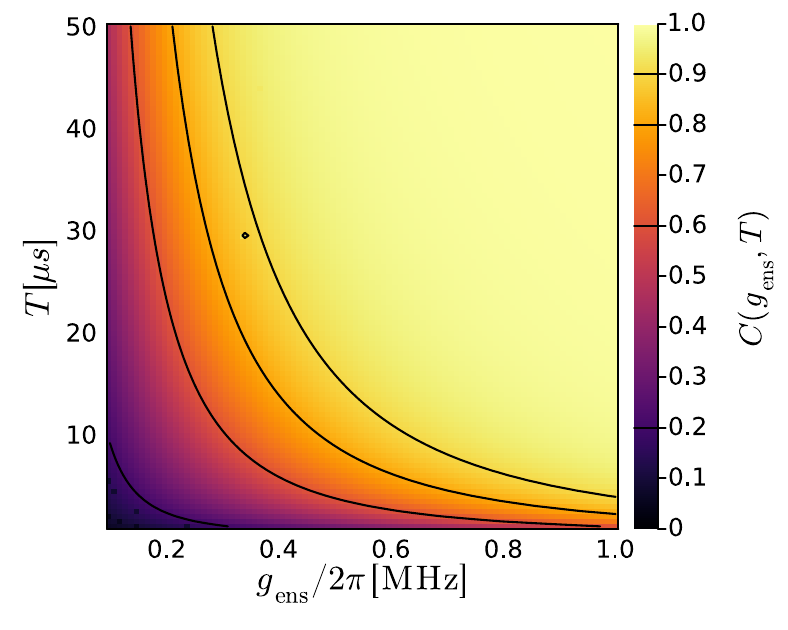}
  \caption{Optimal cooperativity $C=4 g_\text{ens}^2/(\kappa w)$ as a function of $g_\text{ens}$ and $T$ at $w/2\pi= 10$ MHz.
    \label{fig:C_of_g_T}}
\end{figure}
% figure* with two subfigures of Time_2_P_s_heatmap_optim and Speedup_heatmap_optim 
\begin{figure*}
  \centering
  %two subfigures
  \begin{subfigure}[t]{0.45\textwidth}
    \includegraphics[width=8cm]{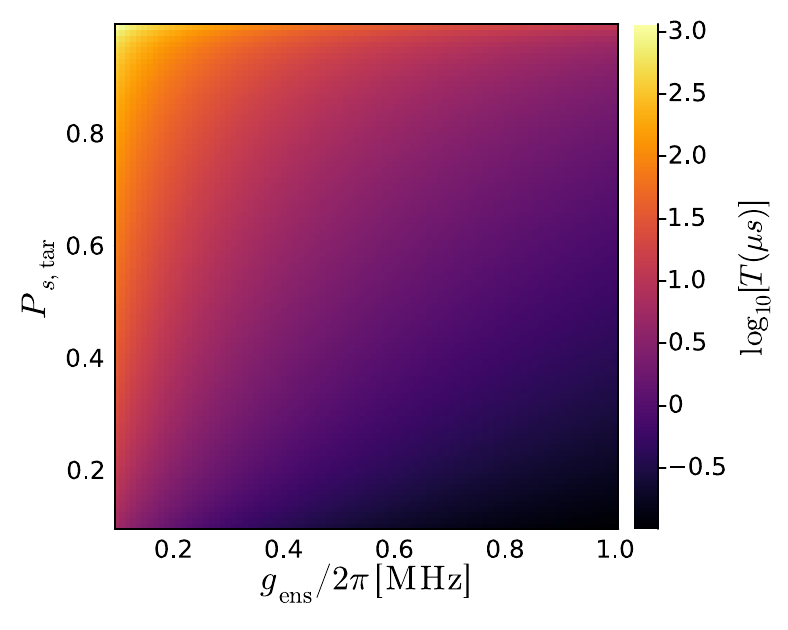}
    \caption{Time to reach target $P_{s, \text{tar}}$ for optimised pulse shapes.}
    \label{fig:Time_2_P_s_heatmap_optim}
  \end{subfigure}
  \hspace{1cm}
  \begin{subfigure}[t]{0.45\textwidth}
    \includegraphics[width=8cm]{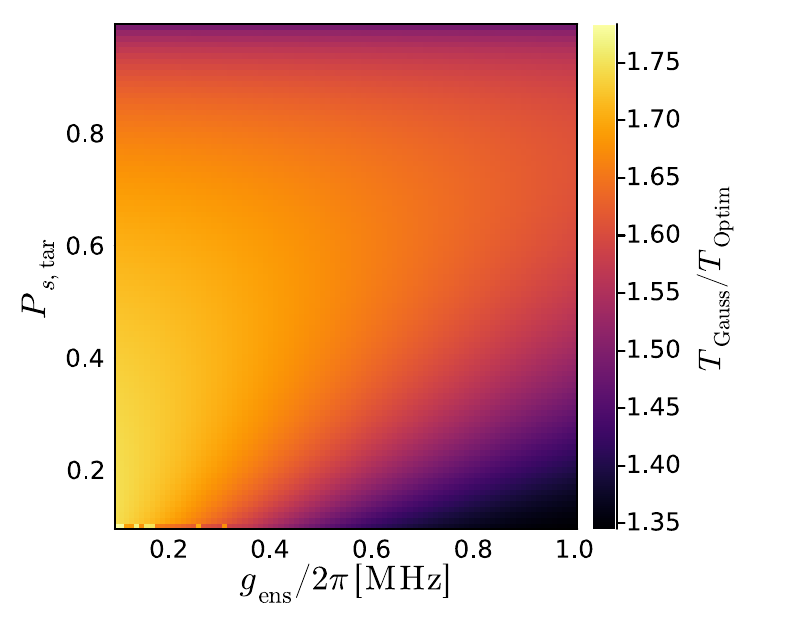}
    \caption{Speedup of optimised pulse shapes.}
    \label{fig:Speedup_of_g}
  \end{subfigure}
  \caption{(a) Logarithm of the time required to reach a target $P_{s, \text{tar}}$ value as a function of $g_\text{ens}$ and (b) the corresponding speedup of optimised pulse shapes over Gaussians using $w/2\pi= 10$ MHz and individually optimised $\kappa$ values.
    \label{fig:Time_2_P_s_heatmap_and_Speedup_of_g}}
\end{figure*}

{\it Results}. In the previous section, we have demonstrated that the Gaussian pulse shapes under the condition of long pulse duration can achieve almost perfect probabilities $P_s$. Now, we compare the bottom panel in Fig.~\ref{fig:Gauss} with the results of the optimised pulse shapes. In Fig.~\ref{fig:P_s_of_kappa_g_contour}, it is demonstrated that our optimisation approach yields a better performance. We solve the Gaussian case, by expanding the function into the first eleven basis functions, i.e., $N_b=5$ in Eq.~\eqref{eq:basisf}. This is a sufficiently accurate approximation, granting a relative error smaller than $1\%$ (and in tests comparing $N_b=5$ to $N_b=10$ giving 4 significant digits) and allows us to reuse the methodology based on $\mathbf{c}$. An optimised pulse is presented in Fig.~\ref{fig:optimal_pulse} and compared
to its Gaussian counterpart of Eq.~\eqref{eq:f1stepGauss} with $\kappa=\kappa_g$. Initially, we run our optimisations with $N_b=5$. In doing so, we find for all optimisations, that optimal pulses depend only on the constant ($j=0$), the first cosine term ($j=1$), and the first sine term ($j=N_b+1$) in Eq.~\eqref{eq:basisf}. Thus, we find an optimal pulse

\begin{equation}
    f(t) = A+ B \sin\left( \frac{\pi t}{T} +\phi \right), \label{eq:opt_pulse}
\end{equation}
where the variables $A$ and $B$ are fixed by the normalisation in Eq.~\eqref{eq:normalization}, while $\phi$ is numerically determined.
This has motivated a cut-off for the optimisations using only $N_b=1$, because increasing $N_b$ to five makes only an order of $~10^{-9}$ difference in $P_s$, comparable to our numerical integration error.

In Fig.~\ref{fig:C_of_g_T}, we show the optimal cooperativity $C=4 g_\text{ens}^2/(\kappa w)$ as a function of $g_\text{ens}$ and $T$. This reproduces previous results of Ref. \cite{Afzelius} suggesting, that for large times $T$ and coupling strengths $g_\text{ens}$, the ideal cooperativity takes the value $1.0$. However, for current experimental setups \cite{Wen2022} $g_\text{ens}=350$ kHz and $T$ is at most $10 \mu$s, which suggests that optimal probability $P_s$ is achieved for values of $C$ between $0.6$ and $0.7$. It is worth noting that for given values of $g_\text{ens}$, $T$ and optimised $\kappa$ $P_s$ may not be equal to one. Finally,
we focus on a different perspective: for optimised values of $\kappa$ and a target probability, how does the pulse duration $T$ depend on given values of $g_\text{ens}$. Our findings in Fig.~\ref{fig:Time_2_P_s_heatmap_and_Speedup_of_g} demonstrate if $g_\text{ens}$ is not large enough then this can be only compensated by larger values of $T$. As higher is the target probability, more demanding are the conditions for both $g_\text{ens}$ and $T$. We have also compared in Fig.~\ref{fig:Time_2_P_s_heatmap_and_Speedup_of_g} the required time for a Gaussian and an optimised pulse shape.
We find an approximately $1.7$-fold speedup over the Gaussian pulses for the experimentally relevant regime with $g_\text{ens}/2\pi \in [0.2, 0.4]\,\text{MHz}$, and limited dependence on $P_s$. For the highest target values of $P_s$, the speedup decreases slightly, this might be an artefact due to limitations in the root finding precision for exceedingly long pulse durations $T$. 

\section{Summary and conclusions}
\label{VI}

In the context of optimal photon storage in a spin ensemble-based quantum memory, we have presented a minimal model capable of describing the physics of such a system. We have shown the analytical solutions of the 
time evolution and the required approximations. Based on these results we have proposed a two-step and several one-step protocols. By using parameters corresponding to current experiments, we have found several pros and cons of the proposed protocols. The two-step protocol can achieve a good absorption of the photon, but this happens for large values of ensemble-coupling $g_{\text{ens}}$. The one-step protocols can achieve good absorption probabilities for lower values of $g_{\text{ens}}$ and in the case of Gaussian pulses with a broad temporal profile one can obtain perfect absorption. The maximum is reached for certain values of $\kappa$ and $g_{\text{ens}}$. Furthermore, the one-step protocols are characterised by a repulsive character, i.e., a too large ensemble-coupling prevents the photon from being absorbed. In general, as one would expect, we have observed that low values of the total cavity decay and the linewidth of the inhomogeneous broadening $w$ improve the success of the protocols. Finally, numerical optimisations of the pulse shapes have resulted in shorter protocol times and higher absorption probabilities of the photon. This has also motivated a new pulse shape, a half-period sine pulse shifted upwards (Fig.~\ref{fig:optimal_pulse}), which can speed up absorption. Throughout the whole manuscript, we have indicated that the excitation probabilities depend on the ratios $\alpha$ and $\alpha r$, and almost perfect excitations are possible if the internal cavity loss $\kappa_i$ is much smaller than $\kappa_e$. Such a condition implies that $\alpha=\sqrt{\kappa_e/\kappa} \approx 1$ and if in addition $2|\Delta_{cs}| \ll \kappa$ then $r \approx 1$. These conditions depend only on the experimental platform and are independent of the optimisation procedures presented here.

In conclusion, our theoretical search for optimal storage of a photon serves as a prerequisite for more advanced tasks, such as storing quantum states in a long-time memory or reobtaining the state, for example, by using the Hahn echo \cite{Hahn}. The simulations and optimisations of these tasks require a master equation approach and the system's state is described by a density operator. The difficulty lies in the fact that the state of the spin ensemble after the absorption is highly entangled, which has to be protected from collective spontaneous emission \cite{Tanas} and/or dephasing \cite{Carnio}. Furthermore, the state of the spin ensemble has to be refocused via $\pi$ pulses, which are implemented with external driving fields of many photons. This requires other mathematical approaches and together with the results of this manuscript, they can provide a complete description of inhomogeneous spin memories.

\begin{acknowledgments}
This work was supported by AIDAS-AI, Data Analytics
and Scalable Simulation, which is a Joint Virtual Laboratory
gathering the Forschungszentrum J\"ulich and the French
Alternative Energies and Atomic Energy Commission, as well as by HORIZON-CL4-2022-QUANTUM-01-SGA Project under Grant 101113946 OpenSuperQPlus100, by Germany’s Excellence Strategy – Cluster of Excellence Matter and Light for Quantum Computing (ML4Q) EXC 2004/1 – 390534769, by the Helmholtz Validation Fund project “Qruise” (HVF-00096), and by the German Federal Ministry of Research (BMBF) under the project SPINNING (No. 13N16210).
\end{acknowledgments}

%%%% Appendix
\appendix

\section{Gaussian distribution of the broadened spin ensemble}
\label{AppendixA}

In the main text, we have mainly considered a Cauchy-Lorentz distribution and now we show the details concerning a Gaussian probability distribution in Eq.~\eqref{eq:Gaussd}.
Then, we have
\begin{eqnarray}
 \Delta_N=\sum^N_{i=1} \frac{\Delta_i}{2} \rightarrow \int^\infty_{-\infty} p_1(\Delta) \frac{\Delta}{2}\, d\Delta=0
\end{eqnarray}
and
\begin{eqnarray}
 &&\sum^N_{i=1} e^{i (\Delta_N -\Delta_i) (t-t')} \rightarrow \int^\infty_{-\infty}\,
 p_1(\Delta)   e^{-i \Delta (t-t')} d\Delta \nonumber \\
 &&= e^{-w^2 (t-t')^2/2}.
\end{eqnarray}

Thus, Eq.~\eqref{eq:cthree} with $\Delta_{cs}=0$ reads
\begin{equation}
 \dot{\Psi}_c=-g^2_{\text{ens}} \int^t_0  e^{-w^2 (t-t')^2/2} \Psi_c(t')\,dt'+\alpha f(t) - \frac{\kappa}{2}\Psi_c.
 \label{A1}
\end{equation}
A general solution to this equation can be obtained with the
help of the Laplace transformation
\begin{equation}
\Psi_c(z)=\int^\infty_0 \Psi_c(t) e^{-zt}\,dt. 
\end{equation}
We use the properties of the Laplace transform on Eq.~\eqref{A1} to obtain
\begin{eqnarray}
-\Psi_c(0)+z\Psi_c(z)&=&-g^2_{\text{ens}} \frac{\sqrt{\frac{\pi }{2}} e^{\frac{z^2}{2 w^2}} \operatorname{erfc}\left(\frac{z}{\sqrt{2} w}\right)}{w} \Psi_c(z) \nonumber \\
&+&\alpha f(z)-\frac{\kappa}{2}\Psi_c(z), \label{eq:A6}
\end{eqnarray}
where $\operatorname{erfc}$ is the complementary error function. The solution is
\begin{equation}
\Psi_c(z)=\frac{\Psi_c(0)}{P(z)}+ \alpha \frac{f(z)}{P(z)}
\end{equation}
with
\begin{equation}
P(z)=z+\frac{\kappa}{2}+g^2_{\text{ens}} \frac{\sqrt{\frac{\pi }{2}} e^{\frac{z^2}{2 w^2}} \operatorname{erfc}\left(\frac{z}{\sqrt{2} w}\right)}{w}.\label{eq:A8}
\end{equation}
To evaluate the inverse Laplace transform, one needs to solve $P(z)=0$, which is a transcendental equation
and has only approximate numerical solutions. In fact, all of them have to be found to obtain the solution to Eq.~\eqref{A1}. 
Instead of searching on the whole complex plane, one can approximate the Gaussian distribution in \eqref{eq:Gaussd} as
\begin{equation}
    p_1(\Delta) \approx \frac{1}{\sqrt{2 \pi} w } \sum^{M_1}_{i=1} \frac{a_i}{b_i+\Delta^2},
\end{equation}
where $a_i,b_i \in \mathbb{R}$ and $b_i \geqslant 0$ for all $i$.

We implement the gradient descent method \cite{boyd} by using $M_2$ points to discretise $p_1(\Delta)$. At each point 
$\Delta_j$, we compute the squared distance between the actual function value and the estimated function value. Finally, we sum up all terms to obtain the cost function
\begin{equation}
    C= \sum^{M_2}_{j=1} \left(p_1(\Delta_j) - \frac{1}{\sqrt{2 \pi} w }\sum^{M_1}_{i=1} \frac{a_i}{b_i+\Delta^2_j} \right)^2.
\end{equation}
Now, we use the update equations:
\begin{equation}
a_i-\gamma \pdv {C}{a_i} \to a_i \quad \text{and} \quad 
b_i-\gamma \pdv {C}{b_i} \to b_i,
\end{equation}
where $\gamma$ is the learning rate. This approach leads to 
a different Laplace transformed equation:
\begin{eqnarray}
-\Psi_c(0)+z\Psi_c(z)&=&-g^2_{\text{ens}} 
\sum^{M_1}_{i=1} \frac{a_i}{w} \sqrt{\frac{\pi}{2 b_i}} \frac{1}{\sqrt{b_i}+z}\Psi_c(z) \nonumber \\
&+&\alpha f(z)-\frac{\kappa}{2}\Psi_c(z),
\end{eqnarray}
see Eq.~\eqref{eq:A6} for a comparison. In place of \eqref{eq:A8}, we have
\begin{equation}
P(z)=z+\frac{\kappa}{2}+g^2_{\text{ens}} \sum^{M_1}_{i=1} \frac{a_i}{w} \sqrt{\frac{\pi}{2 b_i}} \frac{1}{\sqrt{b_i}+z}
\end{equation}
and $P(z)=0$ results in a problem, where the roots of a $(M_1+1)$th degree polynomial have to be determined. Numerically, this is a simpler task than solving a transcendental equation, because we have to find exactly 
$(M_1+1)$ roots.

\section{Constraint on the pulse shapes in one-step protocols}
\label{AppendixB}

The pulse shape $f_e(t)$ acts as an external drive and depends only on the properties of the external field and their couplings to the single-mode field inside the cavity (see Eq.~\eqref{eq:initial}). The probability amplitude $\Psi_c$ of the single-mode in the cavity is governed by Hamiltonian dynamics, however, after the Weisskopf-Wigner approximation is subject to decay. In this approximated theory, the question is what are the properties of $f_e(t)$ such that $\Psi_c$ remains a probability amplitude? This issue is not related to the spin ensemble and its interaction with the single-mode field or the internal decay rate $\kappa_i$ of the cavity. Therefore, we start with thereduced differential equations 
\begin{eqnarray}
  \dot{\Psi}_c&=&-i \sum_{j \in L} \kappa^*_j \Psi^{(j)}_E,  \\
  \dot{\Psi}^{(j)}_E&=&- i \delta_j \Psi^{(j)}_E-i\kappa_j \Psi_c. \label{eq:psi_e_diff}
\end{eqnarray}
For the initial conditions, similarly to the main text, we assume that the excitation is in the external field
\begin{equation}
\label{eq:unitprob}
 \Psi_c (0)=0, \quad \text{and} \quad  \sum_{j \in L} |\Psi^{(j)}_E(0)|^2=1. 
 \end{equation}
We have already solved this for $\Psi_c (t)$ in Eq.~\eqref{eq:cone}, which reads now
\begin{eqnarray}
 &&\dot{\Psi}_c=\underbrace{-i \sum_{j \in L} \kappa^*_j \Psi^{(j)}_E(0) e^{-i \delta_j t}}_{=f_e(t)} \label{eq:useft}\\
 &&- \sum_{j \in L} |\kappa_j|^2 
 \int^t_0 e^{-i \delta_j (t-t')} \Psi_c(t')\, dt'. \nonumber
\end{eqnarray}
Now, as it is explained at Eq.~\eqref{eq:sumtoint}, we replace the summation over the modes with an integral to have
\begin{eqnarray}
\label{eq:B5}
&&\sum_{j \in L} |\kappa_j|^2 
 \int^t_0 e^{-i \delta_j (t-t')} \Psi_c(t')\, dt'  \\
 &&= \int^\infty_{-\infty} d \delta\, \tilde \rho_E(\delta)|\kappa_e(\delta)|^2 \int^t_0 e^{-i \delta (t-t')} \Psi_c(t')\, dt', \nonumber
\end{eqnarray}
where we have introduced the density of states $\rho_E(\delta)$ instead of $\rho_E(\mathbf{k})$. This step allows us to get quicker to the property of $f_e(t)$ than using the integration over $\mathbf{k}$. 

In the next step, we assume 
\begin{equation}
\label{eq:B6}
 \tilde \rho_E(\delta)|\kappa_e(\delta)|^2 \approx \frac{\kappa_e}{2 \pi},   
\end{equation}
i.e., this quantity varies little as a function of $\delta$. The integral
\begin{eqnarray}
  &&\int^\infty_{-\infty}\, d\delta \,e^{-i \delta (t-t')}=
  \lim_{\epsilon \to 0} \int^\infty_{0}\, d\delta \,e^{-i \delta (t-t')-\epsilon \delta} \nonumber \\
  &&+\lim_{\epsilon \to 0} \int^0_{-\infty}\, d\delta \,e^{-i \delta (t-t')+\epsilon \delta}=
  2 \pi \delta(t-t'),  
\end{eqnarray}
where $\epsilon>0$ and $\delta(t)$ is the Dirac delta function. Here, we have used the following representation of the Dirac delta function:
\begin{equation}
\label{eq:DDrepr}
  \delta (t-t')= \lim_{\epsilon \to 0} \frac{1}{\pi} \frac{\epsilon}{(t-t')^2+\epsilon^2} 
\end{equation}
Then, \eqref{eq:B5} reads
\begin{eqnarray}
 \frac{\kappa_e}{2 \pi} 2 \pi  \int^t_0 \delta(t-t') \Psi_c(t')\, dt' =\frac{\kappa_e}{2} \Psi_c(t),
\end{eqnarray}
because the integration from $0$ to $t$ covers
only half of the function in \eqref{eq:DDrepr}, which is symmetrical about the $t=t'$ vertical line.

We shall now extend this approximation to the external drive $f_e(t)$. According to Eqs. \eqref{eq:unitprob} and \eqref{eq:useft} we have
\begin{eqnarray}
 1&=& \sum_{j \in L} |\Psi^{(j)}_E(0)|^2 = \int^\infty_{-\infty} d \delta\, \tilde \rho_E(\delta) 
 |\Psi_E(\delta, 0)|^2 \nonumber \\
 &=& \int^\infty_{-\infty} d \delta\,   
 |\tilde \Psi_E(\delta, 0)|^2, \label{eq:B12}
\end{eqnarray}
where we have absorbed the square root of the positive function $\tilde \rho_E(\delta)$ into $\tilde \Psi_E(\delta, 0)$, and
\begin{eqnarray}
&&f_e(t)=-i \sum_{j \in L} \kappa^*_j \Psi^{(j)}_E(0) e^{-i \delta_j t} \\
&&= -i \int^\infty_{-\infty} d \delta\, \sqrt{\tilde \rho_E(\delta)}\kappa^*_e(\delta) \tilde \Psi_E(\delta, 0) e^{-i \delta t}. \label{eq:B11}
\end{eqnarray}
We need to assume that $\kappa^*_j$ for all $j$ and $\kappa^*_e(\delta)$ are real to introduce in \eqref{eq:B11} the following relation
\begin{equation}
    \sqrt{\tilde \rho_E(\delta)} \kappa_e(\delta)=\frac{\sqrt{\kappa_e}}{\sqrt{2 \pi}}
\end{equation}
based on \eqref{eq:B6}. Hence,
\begin{equation}
 f_e(t)=-i \sqrt{\kappa_e} \underbrace{\frac{1}{\sqrt{2 \pi}} \int^\infty_{-\infty} d \delta\, \tilde \Psi_E(\delta, 0) e^{-i \delta t}}_{=\tilde \Psi_E(t, 0)},   
\end{equation}
where $\tilde \Psi_E(t,0)$ is the Fourier transform of $\tilde \Psi_E(\delta,0)$. Then, we have
\begin{equation}
 \int^\infty_{-\infty} |f_e(t)|^2 \,dt = \kappa_e \int^\infty_{-\infty} |\tilde \Psi_E(t,0)|^2 \,dt.
\end{equation}
Now, we use Eq.~\eqref{eq:B12} and Plancherel's theorem \cite{Yosida}, which states that the Fourier transform map is an isometry with respect to the $L^2$ norm:
\begin{equation}
 \int^\infty_{-\infty} |\tilde \Psi_E(t,0)|^2 \,dt = \int^\infty_{-\infty} |\tilde \Psi_E(\delta,0)|^2 \, d \delta =1.
\end{equation}
It is immediate
\begin{equation}
 \int^\infty_{-\infty} |f_e(t)|^2 \,dt=\kappa_e.
\end{equation}

\section{Optimisation and Integration} \label{sec:optimization_apendix}

As we have described in Sec. \ref{sec:optnum}, the basis functions are:
\begin{eqnarray}
  f_j(t) = \begin{cases}
    \cos\left(\pi \frac{j t}{T} \right)       
    & \text{for } 0\leqslant j \leqslant N_b, \\
    \sin\left(\pi \frac{(j-N_b) t}{T} \right) & \text{for } N_b<j \leqslant 2N_b.
  \end{cases}
\end{eqnarray}
 We define the coefficient vector $\mathbf{c}=(c_0, c_1, \dots, c_{N_b})^\mathrm{T}$ ($\mathrm{T}$ denotes the transposition). A computationally quicker approach to calculate values $P_s\left[f(t)\right]$ via Eq.~\eqref{eq:inneruse} is to expand the equation in the basis functions and cache the respective integrals for each term. This avoids recomputing them throughout the optimisation process. We have
\begin{eqnarray}
  P_s&=&g^2_{\text{ens}} \int^\infty_{-\infty} d\Delta\, p_1(\Delta)  \left | \int^T_0\, dt' f(t') h(t',T, \Delta) \right|^2 \nonumber\\
  &=&\sum_{i,j=0}^{2N_b} c_i c_j^{*} \underbrace{\int^\infty_{-\infty} d\Delta\, p_1(\Delta) I_i(\Delta) I_j^{*}(\Delta)}_{=P_{ij}}, \label{eq:inneruse_basis}
\end{eqnarray}
where we have used $I_i(\Delta) = \int^T_0\, dt' f_i(t') h(t',T, \Delta)$ for the inner integral terms. In order to solve the inner integral from Eq.~\eqref{eq:inneruse} for the $j^\text{th}$ basis function in \eqref{eq:basisf},
we define the terms $m(t, T)= e^{-\frac{\kappa + w}{4} (T-t')} \cosh \left[\frac{\varpi'}{4} (T-t')\right] $ and
$n(t, T) = e^{-\frac{\kappa + w}{4} (T-t')} \sinh \left[\frac{\varpi'}{4} (T-t')\right]$, so that
\begin{eqnarray}
  I_j(\Delta) &=& \int^T_0\, dt' f_j(t') h(t',T, \Delta) \nonumber \\
  &=& A(\Delta) \underbrace{\int^T_0\, dt' f_j(t') m(t', T)}_{=M_j(T)}  \nonumber\\
  &+& B(\Delta) \underbrace{\int^T_0\, dt' f_j(t') n(t', T)}_{=N_j(T)}  \nonumber\\
  &-& A(\Delta) \underbrace{\int^T_0\, dt' f_j(t') e^{-i\Delta (T-t')}}_{=D_j(T,\Delta)}. \label{eq:innerint}
\end{eqnarray}

Let us first solve the final integral in Eq.~\eqref{eq:innerint}.
We find for the sine and cosine basis functions with $0<j\leqslant 2N_b$
\begin{eqnarray}
  D_j(T, \Delta) &=& \int^T_0\, dt' f_j(t') e^{-i\Delta (T-t')} \nonumber\\
  &=& \begin{cases}
    \int^T_0\, dt' \sin\left(\pi \frac{j t}{T} \right) e^{-i\Delta (T-t')} & \text{for } j \leqslant N_b, \\
    \int^T_0\, dt' \cos\left(\pi \frac{j t}{T} \right) e^{-i\Delta (T-t')} & \text{for } j > N_b,
  \end{cases} \nonumber \\
  &=& \begin{cases}
    \frac{\pi T j \left[\left(-1\right)^{j}  -  e^{- i T \Delta}\right]}{T^{2} \Delta^{2} - \pi^{2} j^{2}}      & \text{for } j \leqslant N_b, \\
    \frac{i \Delta T^2 \left[\left(-1\right)^{j}  -  e^{- i T \Delta}\right]}{T^{2} \Delta^{2} - \pi^{2} j^{2}} & \text{for } j > N_b.
  \end{cases}
\end{eqnarray}
It should be noted, that the divergences at $\Delta = \pm \frac{\pi j}{T}$ are resolvable via L'Hôpital's rule, so that
\begin{eqnarray}
  D_j(T, \Delta=\pm \frac{\pi j}{T}) =  \begin{cases}
    \frac{i \pi j (-1)^k}{2 \Delta} & \text{for } j = k \leqslant N_b,   \\
    \frac{-T (-1)^{k}}{2 }          & \text{for } j - N_b = k > 0.
  \end{cases} \nonumber \\
\end{eqnarray}
We can expand the sine, cosine, hyperbolic sine and hyperbolic cosine terms in the remaining integrals $M_j(T)$ and $N_j(T)$ in \eqref{eq:innerint} into exponentials, which we can integrate analytically.
Every term will be of the form below, where we use $\pm $ for the sign of the exponentials of the hyperbolic sine and hyperbolic cosine terms and $\mp$ for the sign of the exponentials of the sine and cosine terms
\begin{eqnarray}
  E(\pm, \mp) &=& \int^T_0\, dt' e^{-\frac{\kappa + w}{4} (T-t')} e^{\pm \frac{\varpi'}{4} (T-t')} e^{\mp i \pi \frac{j t'}{T}} \nonumber \\
  &=& \frac{1}{\frac{ \pm \varpi' -\kappa - w}{4} \mp i \frac{\pi j}{T}} \left[ e^{\frac{ \pm \varpi' -\kappa - w}{4} T} - (-1)^j \right]. \nonumber \\
\end{eqnarray}
This allows us to represent all three integrals analytically. Lastly, we keep in mind, that $B(\Delta)N_j(T)$ contains a singularity at $\varpi' = 0$, that would make the numerical integration unstable.
This singularity can be removed via L'Hôpital's rule, but in practice it is avoided by not evaluating at $\varpi' = 0$ and furthermore, this occurs rarely due to the investigated parameter regions.

The $\Delta$ integration for $P_{ij}$ in \eqref{eq:inneruse_basis} is performed numerically for all $i$ and $j$ using the adaptive Gauss-Kronrod quadrature in the  Julia library \textit{QuadGK.jl} \cite{quadgk} with a relative precision of $10^{-10}$ and the default G7-K15 rule.
We reduce the integral bounds at infinity to the finite $[-\Delta_\text{max}, \Delta_\text{max}]$, with $\Delta_\text{max}=100$ MHz which guarantees at least 5 significant digits due to the integral arguments being at least of order $\mathcal{O}\left(\frac{1}{\Delta^4}\right)$.
From the cached $P_{ij}$ we can then calculate $P_s$ for any pulse shape $f(t)$, via matrix-vector multiplication $P_s=\mathbf{c}^{\mathrm{T}} \hat{P} \mathbf{c}$, where $\hat{P}$ is the matrix with entries $P_{ij}$. 

The coefficients are constrained due to Eq.~\eqref{eq:normalization} (see also Appendix \ref{AppendixB}). We expand again the integral into the basis functions
\begin{eqnarray}
  \int^T_0\, dt' |f(t')|^2 &=& \sum_{i,j=1}^{2N_b} c_i c_j \underbrace{\int_0^T dt' f_i(t') f_j(t')}_{=F_{ij}} \nonumber \\
  &=& \mathbf{c}^{\mathrm{T}} \hat{F} \mathbf{c} \overset{!}{=} \kappa, \label{eq:coeff_constraint}
\end{eqnarray}
where the entries of the constraint matrix $\hat{F}$ are calculated analytically, which are given by
\begin{eqnarray} 
  F_{j,j} &=& \begin{cases}
    T           & \text{for } j = 0,     \\
    \frac{T}{2} & \text{for } j \neq 0,
  \end{cases}\\
  F_{j, N_b+k} &=& F_{N_b+k, j} = \begin{cases}
    2 T \frac{1-(-1)^k}{\pi k}               & \text{for } j = 0 < k,    \\
    2 T j \frac{1-(-1)^{j+k}}{\pi (j^2-k^2)} & \text{for } j \neq k > 0, \\
    0                                        & \text{for } j = k > 0.
  \end{cases} \nonumber
\end{eqnarray}
We normalise $\mathbf{c}$ before calculating $P_s$ to satisfy the constraint in \eqref{eq:coeff_constraint} \footnote{Constrained optimisation sometimes failed to keep the constraint satisfied, so we opted for this approach.}. In the optimisation, we rescale any input coefficient vector $\mathbf{c}$, which violates the constraint, by the factor $\sqrt{\frac{\kappa}{\mathbf{c}^{\mathrm{T}} \hat{F} \mathbf{c}}}$.

{\it Optimisation procedure}. Having reduced both the problem of finding the probability $P_s$ and similarly the constraint calculation to matrix-vector multiplications of cached integrals $\hat{P}$, $\hat{F}$ and basis function coefficients in $\mathbf{c}$, we can now efficiently optimise the pulse shapes $f(t)$ for any given set of parameters $w$, $g_{\text{ens}}$ and $\kappa$. We restate the optimisation function Eq.~\eqref{eq:f_opt_prob} for the optimisation of the constrained basis function coefficients in $\mathbf{c}$ to
\begin{equation}
    \mathbf{c}_\text{opt} = \underset{\mathbf{c}}{\operatorname{argmax}}\, P_s(\mathbf{c}, T, w, g_\text{ens}, \kappa). \label{eq:f_opt_prob2}
\end{equation}
We solve this for the optimal coefficients using the Broyden–Fletcher–Goldfarb–Shanno (BFGS) method \cite{Shanno1970} via the implementation in the Julia library \textit{Optim.jl} \cite{Optim.jl-2018}, where we also use automatic (forward) differentiation in \textit{ForwardDiff.jl} \cite{RevelsLubinPapamarkou2016}.

In practice \cite{Wen2022}, the parameter $\kappa$ can be changed, whereas $g_\text{ens}$ is given by the inherent dipole coupling strengths of the spins.
To exploit this degree of freedom, we have also developed a method to find both the optimal $\kappa$ and optimal pulse shape $f(t)$ for a given set of parameters $g_{\text{ens}}$, $w$ and $T$.
In principle, we could define both parameters as optimisation variables, but since the \textit{QuadGK.jl} library uses an adaptive integration approach that cannot be automatically differentiated,
we opted for a nested optimisation approach, constructed from an outer optimisation, optimising $\kappa$, and an inner optimisation, optimising $f(t)$ for every outer iteration for the current value of $\kappa$.
Whereas the inner optimisation for $\mathbf{c}_\text{opt}(\kappa)$ uses the aforementioned BFGS method and Eq.~\eqref{eq:f_opt_prob2}, the outer optimisation uses the Newton's root-finding method with finite differences, which is also implemented in \textit{Optim.jl}. This yields 
\begin{eqnarray}
    \kappa_\text{opt} &=& \underset{\kappa}{\operatorname{argmax}}\, P_s(\mathbf{c}_\text{opt}(\kappa), T, w, g_\text{ens}, \kappa).\label{eq:kappa_opt_prob2}
\end{eqnarray}
Finally, we determine the minimum pulse durations $T_\text{min}$ to achieve a target probability $P_{s, \text{tar}}$. The probability increases monotonically with $T$, so we can find the minimum duration via root-finding of  Eq.~\eqref{eq:T_min_prob}. We employ Newton's method for optimisation of the Julia library \textit{Roots.jl} \cite{Roots.jl}, where the gradient is calculated with finite differences.\\

%%%% Bibliography
\bibliography{manuscript_revision}

\end{document}